%% file: paper.tex
\documentclass[11pt]{article}

\usepackage{verbatim}
\usepackage{amsmath}
\usepackage{amsfonts}
\usepackage{amssymb}
\usepackage{latexsym}
\usepackage{graphics}
\usepackage{lscape} 
\usepackage{epsfig}
\usepackage{color}

\input{texdefs}

\begin{document}

\input{title}

\setcounter{page}{1}

\input{intro}

\input{setup}

\input{oneloop}

\input{twoloop}

\input{examples}

\appendix 
\def\theequation{\thesection.\arabic{equation}} 

\setcounter{equation}{0}
\input{1Lintegrals}

\setcounter{equation}{0}
\input{2Lintegrals}

\bibliographystyle{paper}
{\small
\bibliography{paper}
}

\end{document}

%% file: texdefs.tex
\renewcommand{\d}{\mathrm{d}}

\newcommand{\captn}[1]{\vspace{-3ex}\caption{\small #1}}

\newcommand{\dd}[2][]{\ensuremath{\frac{\d #1}{\d #2}}}
\newcommand{\pp}[2][]{\ensuremath{\frac{\partial #1}{\partial #2}}}

\newcommand{\dga}{{\dot{\alpha}}}

\newcommand{\ga}{\alpha}

\renewcommand{\gg}{\gamma}
\newcommand{\gd}{\delta}
\renewcommand{\ge}{\epsilon}

\newcommand{\gf}{\phi}

\newcommand{\gx}{\xi}
\newcommand{\gm}{\mu}

\newcommand{\gk}{\kappa}
\newcommand{\gl}{\lambda}

\newcommand{\gth}{\theta}

\newcommand{\gs}{\sigma}

\newcommand{\gz}{\zeta}
\newcommand{\gp}{\pi}

\newcommand{\gch}{\chi}
\newcommand{\gG}{\Gamma}
\newcommand{\gD}{\Delta}
\newcommand{\gF}{\Phi}

\newcommand{\gL}{\Lambda}
\newcommand{\gS}{\Sigma}
\newcommand{\gTh}{\Theta}

\newcommand{\gPs}{\Psi}

\newcommand{\cM}{{\cal M}}
\newcommand{\cN}{{\cal N}}

\newcommand{\cW}{{\cal W}}

\newcommand{\ua}{{\underline a}}
\newcommand{\ub}{{\underline b}}
\newcommand{\uc}{{\underline c}}
\newcommand{\ud}{{\underline d}}

\newcommand{\Tr}{\mbox{Tr}}
\newcommand{\tr}{\text{tr}}

\newcommand{\Id}{\text{\small 1}\hspace{-3.5pt}\text{1}}

\newcommand{\ra}{\rightarrow}

\newcommand{\der}{\partial}

\newcommand{\inv}{^{-1}}
\newcommand{\dsp}{\displaystyle}

\newcommand{\ubar}[1]{{\underline{#1}}}

\newcommand{\labl}[1]{\label{#1}}
\newcommand{\Kh}{K\"{a}hler}
\newcommand{\beq}{\begin{equation}}
\newcommand{\eeq}{\end{equation}}
\newcommand{\barr}{\begin{array}}
\newcommand{\earr}{\end{array}}
\newcommand{\equ}[1]{\begin{gather} #1 \end{gather}}
\newcommand{\equa}[1]{\begin{align} #1 \end{align}}

\newcommand{\tabu}[2]{\begin{tabular}{#1} #2 \end{tabular}}
\newcommand{\arry}[2]{\begin{array}{#1} #2 \end{array}}

\newcommand{\pmtrx}[1]{\begin{pmatrix} #1 \end{pmatrix}}

\newcommand{\non}{\nonumber}
\newcounter{oldcounter}

\newcommand{\bff}{{\bar f}}

\newcommand{\bm}{{\bar m}}

\newcommand{\bx}{{\bar x}}

\newcommand{\bz}{{\bar z}}

\newcommand{\bC}{{\overline C}}
\newcommand{\bD}{{\overline D}}

\newcommand{\bF}{{\overline F}}

\newcommand{\bI}{{\overline I}}
\newcommand{\bJ}{{\overline J}}

\newcommand{\bW}{{\overline W}}

\newcommand{\bgf}{{\bar\phi}}

\newcommand{\bgm}{{\bar\mu}}

\newcommand{\bgch}{{\bar\chi}}
\newcommand{\bgG}{{\overline\Gamma}}

\newcommand{\bgF}{{\overline\Phi}}

\newcommand{\bgL}{{\overline\Lambda}}

\newcommand{\bgTh}{{\overline\Theta}}

\newcommand{\bcW}{{\overline{\cal W}}}

\newcommand{\Real}{\mathbb{R}}

\newcommand{\ba}[2]{\[\begin{array}{#2}\label{#1}}
\newcommand{\ea}{\end{array}\]}
\newcommand{\be}{\begin{equation}}
\newcommand{\ee}{\end{equation}}
\newcommand{\bea}{\begin{eqnarray}}
\newcommand{\eea}{\end{eqnarray}}

\newcommand{\U}[1]{\mathrm{U(#1)}}

\newcommand{\MSbar}{$\overline{\mbox{MS}}$}
\newcommand{\DRbar}{$\overline{\mbox{DR}}$}
\newcommand{\sfrac}[2]{\mbox{$\frac{#1}{#2}$}}

\newcommand{\bJten}{\bJ{}^a{}_\ua{}^b{}_\ub}
\newcommand{\bJtenG}{\bJ{}^a{}_\ua{}^{IJ}}
\newcommand{\bJtenGG}{\bJ{}^{IJ\, KL}}

\newcommand{\bIten}{\bI{}^a{}_\ua{}^b{}_\ub{}^c{}_\uc}
\newcommand{\bItenG}{\bI{}^a{}_\ua{}^b{}_\ub{}^{IJ}}
\newcommand{\bItenU}{\bI{}^a{}_\ua{}^b{}_\ub{}}
\newcommand{\bItenGG}{\bI{}^a{}_\ua{}^{IJ\,KL}}
\newcommand{\bItenGGG}{\bI{}^{IJ\,KL\, MN}}

%% file: title.tex
\thispagestyle{empty}

\begin{flushright}
IC/2005/101\\
SIAS-CMTP-05-2\\ 
USTC-ICTS-05-11 \\ 
hep-th/0511004
\end{flushright}
\vskip 2 cm
\begin{center}
{\Large {\bf 
Two Loop Effective K\"ahler Potential of (Non--)Renormalizable
\\[2ex] 
Supersymmetric Models 
}
}
\\[0pt]
\vspace{1.23cm}
{\large
{\bf Stefan Groot Nibbelink$^{a,}$\footnote{
{{ {\ {\ {\ E-mail: nibbelin@ustc-sias.cn}}}}}}}, 
{\bf Tino S. Nyawelo$^{b,}$\footnote{
{{ {\ {\ {\ E-mail: tnyawelo@ictp.it}}}}}}},
\bigskip }\\[0pt]
\vspace{0.23cm}
${}^a$ {\it 
Shanghai Institute for Advanced Study, 
University of Science and Technology of China,  \\ 
99 Xiupu Rd, Pudong, Shanghai 201315, P.R.\ China
\\[1ex] 
Interdisciplinary Center for Theoretical Study, 
University of Science and Technology of China,  \\
Hefei, Anhui 230026, P.R.\ China 
 \\}
\vspace{0.23cm}
${}^b$ {\it 
The Abdus Salam ICTP, Strada Costiera 11, 
I--34014 Trieste, Italy 
\\}
\bigskip
\vspace{1.4cm} 
\end{center}
\subsection*{\centering Abstract}

We perform a supergraph computation of the effective K\"ahler
potential at one and two loops for general four dimensional $\cN=1$
supersymmetric theories described by arbitrary K\"ahler potential,
superpotential and gauge kinetic function. We only insist on gauge
invariance of the K\"ahler potential and the superpotential as we
heavily rely on its consequences in the quantum theory. However, we 
do not require gauge invariance  for the gauge kinetic functions, so
that our results can also be applied to anomalous theories that involve
the Green--Schwarz mechanism. We illustrate our two loop results by
considering a few simple models: the (non--)renormalizable
Wess--Zumino model and Super Quantum Electrodynamics.

\newpage 


%% file: intro.tex
\section{Introduction and Summary}
\labl{sc:intro}

%
%

The use of the effective action in quantum field theories has proven
to be a very powerful tool. Following \cite{Jackiw:1974cv} in
particular the quantum effective potential is widely used. (For a
recent application of the computation of the vacuum energy in extra
dimensions see \cite{vonGersdorff:2005ce}.) 
The effective action of a supersymmetric theory up to two derivatives
is encoded in three functions of the chiral multiplets: the \Kh\ 
potential, the superpotential and the gauge kinetic function. The
latter two are required to be holomorphic and are therefore very much
constraint. This is reflected in various non--renormalization theorems
\cite{Grisaru:1979wc} for these functions and lead to results to all
order \cite{Novikov:1982px,Shifman:1986zi}. The works
\cite{Seiberg:1993vc,Seiberg:1994bz,Intriligator:1994jr} show that
even a lot of non--perturbative information can be
obtained.  In particular, in the certain $\cN=2$ theories the full
non--perturbative superpotential has been computed \cite{Seiberg:1994rs}.

The \Kh\ potential on the other hand is only required to be a real
function, and therefore far less constrained. It receives corrections
at all orders in perturbation theory, unless the theory has sufficient
amounts of extended supersymmetries. For example, the \Kh\ potential
of the adjoint chiral multiplet in $\cN=2$ Super Yang--Mills
theory is determined by a prepotential that also fixes the gauge
kinetic function. The computation of the \Kh\ potential at the one
loop level has been performed by many authors
\cite{Grisaru:1996ve,deWit:1996kc,Pickering:1996gt,Buchbinder:1999jw,Brignole:2000kg},
and their results essentially agree with each other.  One loop
corrections the \Kh\ potential have also been computed in the context
of supergravity \cite{Gaillard:1996hs,Gaillard:1994mn,Gaillard:1994sf}.
An extension of such works to extra dimensions can be found in 
\cite{Falkowski:2005fm}.

At the two loop level only some partial results have been
obtained. The two loop \Kh\ potential for the renormalizable
Wess--Zumino model with a cubic superpotential has been computed
\cite{Buchbinder:1994xq,Buchbinder:1994iw,Buchbinder:1994df,Buchbinder:1996cy},
and has been extended to a general non--renormalizable Wess--Zumino
model by some people from the same group
\cite{Buchbinder:2000ve,Petrov:1999qh}.  
The latter result seem suspicious because it does not appear
to be covariant. As far as we are aware no calculation of the two loop
\Kh\ potential has been performed which takes the effect of gauge
interactions into account.

Precisely because for the \Kh\ potential there are no
non--renormalization theorems in general, it is interesting to
understand it quantum corrections at higher orders. The computation of
the effective \Kh\ potential can also be very important for
phenomenological applications, as it encodes the wave function
renormalization of chiral multiplets. The physical masses of 
chiral multiplets can be determined only when the effect of wave
function renormalization is taken into account. Similarly if the gauge
kinetic function is not a pure constant, then the wave function
renormalization of the chiral multiplets that appear in the gauge
kinetic function induce non--holomorphic renormalization of the gauge
couplings. Also when one investigates the effective potential at the
two loop level, knowing the effective \Kh\ potential might prove very
useful. For some of these applications it might be relevant to know
the renormalization of the superpotential at two loops. Here there is
a somewhat surprising result that the superpotential is renormalized,
but only by a finite amount
\cite{Buchbinder:1994iw,Buchbinder:1994df,Buchbinder:1998qv}.  
At the two loop level, for the class of theories we are
considering, the computation of the selfenergy of the chiral multiplets
can easily involve over a hundred diagrams, which is very hard to
manage. As we will see the computation of the effective \Kh\ potential
only requires about eleven supergraphs to be evaluated. The complicated
expressions for the selfenergies are recovered by evaluating the
second mixed derivative of the two loop effective \Kh\ potential.

%
%

In this paper we perform supergraph computations of the \Kh\
potential up to two loops in generic globally supersymmetric
theories. The explicit one and two loop formulae for the effective
\Kh\ potential can be found in sections \ref{sc:oneloop} and
\ref{sc:summary2L}, respectively. The classical theory is not assumed
to be renormalizable,  and we allow for gauge interactions to be
present. The only simplifying assumptions that we are making are that
the classical theory has no more than second derivatives and the
gauged symmetries are realized linearly\footnote{Because it is easy to
  recognize the linear Killing vectors in all our expressions, it
  would be straightforward to generalize them to arbitrary non--linear
  gaugings.}. 
Both our one and two loop results are manifestly target space
diffeomorphism covariant. When we restrict to the ungauged case, 
we see that we obtain the same terms at two loops as  
\cite{Buchbinder:2000ve,Petrov:1999qh}, but with different
coefficients such that the result contains the curvature tensor and
covariant derivatives of the superpotential. This shows that our
results are manifestly covariant. The result of the two loop \Kh\
potential looks surprisingly simple as long as the gauge 
kinetic function is strictly constant.

Apart from the possible applications mentioned above, our results at
the two loop level might be interesting for various applications in
$\cN=2$ theories. In theories with extend supersymmetry the \Kh\
and super--potential are obtained from a single holomorphic
prepotential. Since our results are obtained for generic $\cN=1$
supersymmetric theories, they can be applied in particular to $\cN=2$
theories, and may provide important cross checks on the validity of
the constraints that come from the $\cN=2$ structure. At the one loop
level such computations have been performed
\cite{deWit:1996kc,Pickering:1996gt}. Our results can be used to
undertake a similar analysis at the two loop level.

%
%

Before we close the introduction and summary section, we would like
to explain our methodology in the outline of the paper. In section
\ref{sc:setup} we present the basic starting point of our
calculations: a general $\cN=1$ supersymmetric model based on a \Kh\
manifold with some of its linear isometries gauged. As far as superspace
is concerned we will be using the conventions of Wess and Bagger
\cite{Wess:1992cp}. To compute the one and two loop corrections to the
\Kh\ potential we employ a background field method. Since we are
interested in the effective \Kh\ potential, we assume that the
background consists of chiral multiplet. Since in this paper we are 
focusing on the quantum corrections to the \Kh\ potential only, we will be 
systematically ignoring all derivatives on the background chiral
superfields in the whole project, as these terms constitute higher derivative 
operators. These higher derivative operators are beyond the scoop of the 
present paper. We do not assume that the background is constant, because 
then we are not able to extract the effective \Kh\ potential. In principle, we 
are computing the 1PI effective \Kh\ potential, however, to ensure that 
the relevant propagators do not become too complicated, we make 
use of the background equations of motion.

One consequence of this chiral background is that the gauge
symmetry is generically spontaneously broken. The resulting quadratic
mixing between the quantum vector and chiral multiplets makes
quantization awkward. To overcome this complication we remove this
mixing by using a supersymmetric generalization of the 't Hooft
$R_\gx$ gauge fixing procedure. A consequence of this gauge is that
the ghost action becomes dependent on the background chiral multiplets
and cannot be neglected. Moreover, in this gauge the Goldstone bosons
and the ghosts have equal mass eigenvalues. We will be making a
special choice for the $R_\gx$ gauge fixing parameter to have the
maximal simplification of the super propagators. Because of this 
we loose the explicit dependence on a gauge fixing parameter and
therefore we are not able to trace any possible gauge dependence of
our results.

In section \ref{sc:oneloop} we review the computation of the one
loop \Kh\ potential by summing bubble to explain our procedure to
reduce supergraph to scalar graphs. We use standard supergraph
techniques that can be found in the textbooks
\cite{Gates:1983nr,West:1990tg}. To regularize the theory we employ
the dimensional reduction scheme (\DRbar)
\cite{Siegel:1979wq,Capper:1979ns}.  (Following arguments 
summarized in refs.\ \cite{Jack:1994bn,Jack:1997sr} we assume that for
our purposes the \DRbar\ scheme is fully consistent.)  A well--known
consequence of this scheme is that quadratic divergences do not show
up explicitly. (This will also be the case at the two loop level,
therefore our results cannot be directly compared to those of
\cite{Bagger:1995ay}.) Our one loop results are consistent with
results obtained by previous groups 
\cite{Grisaru:1996ve,deWit:1996kc,Pickering:1996gt,Buchbinder:1999jw,Brignole:2000kg},
as long as we do not consider non--Abelian gauge interactions. The
difference in that case is presumably a consequence of the fact that
those results are obtained in the Landau gauge while we use a
supersymmetric variant of the 't Hooft $R_\gx$ gauge.

The core of our paper is contained in section \ref{sc:twoloop} where
we perform our two loop computation. We first compute a single diagram
of the ``8'' topology which quickly reduces to a product of two
one--loop scalar graphs. After that we turn to the other
nine two--loop supergraphs with the ``$\ominus$'' topology. Some of
them have the property that they partly turn into ``8'' scalar graphs. 
The resulting scalar graphs are evaluated in appendix
\ref{sc:2Lintegrals}. We summarize the results of the contributions to
the \Kh\ potential at two loop in subsection \ref{sc:summary2L}, and
compare with the partial results that exist in the literature on two
loop effective \Kh\ potential calculations.

Since our results are written in terms of some tensor valued
integrals, it is useful to see how the formulae can be applied to some
simple examples. In section \ref{sc:examples} we discuss both the
non--renormalizable Wess--Zumino model and its renormalizable limit. 
Super Quantum Electrodynamics constitutes our second example.

\section*{Acknowledgments}

We would like to thank T.\ Hirayama 
and R.\ Zwicky for pleasant discussions at various stages of this
project. We grateful to A.\ Brignole for communications on the
one loop \Kh\ potential. We have benefitted from correspondence with
C.\ Ford, I.\ Jack and D.R.T.\ Jones concerning the computation of two
loop scalar graphs. We are also grateful to S.J.\ Tyler for pointing
out a missing contribution and some typos in the published version of
this work.


%% file: setup.tex
\section{Setup: $\boldsymbol{\cN=1}$ gauge non-linear sigma model}
\labl{sc:setup} 

We consider a general $\cN=1$ supersymmetric non--linear sigma model
of (anti--)chiral superfields $\gf^a (\bgf_\ua)$ described by a \Kh\
potential $K(\bgf, \gf)$ and a superpotential $W(\gf)\,$.  The indices
$a$  and $\ua$ label the chiral and anti--chiral multiplets,
respectively. Some of the  linear isometries 
$\gd_\ga \gf ~=~ i \ga \, \gf ~=~ i \ga^I T_I \, \gf$ 
(where $\ga_I \in \Real$ are gauge parameters) 
are assumed to be gauged by the introduction of the non--Abelian gauge
vector superfield  $V ~=~ V^I T_I\,.$ The Hermitean generators $T_I$
of this group satisfy the algebra    
\(
[T_I, T_J] ~=~ c^K{}_{IJ} \, T_K\,, 
\)
with purely imaginary structure coefficients $c^K{}_{IJ}\,$. 
The Killing metric is denoted by $\gd_{IJ}\,.$ We write $\tr$ and
$\Tr$ for the traces over the representation of the chiral multiplets
$\gf$ and the adjoint representation, respectively. Gauging is of
course only possible if the \Kh\ potential and the superpotential
are gauge invariant  
\equ{
K(\bgf\, e^{-i\ga}, e^{i \ga} \,\gf) - K(\bgf, \gf) ~=~ 0~,
\qquad 
W(e^{i\ga} \, \gf) ~=~ W(\gf)~. 
\labl{GaugeInv}
}
This will have important consequences for the quantum interactions. 
Unlike the \Kh\ potential or the superpotential, the gauge kinetic
function $f_{IJ}(\gf)$ need not be gauge invariant, as it may play a
role in a Green--Schwarz mechanism to cancel anomalies.

The central task of this paper is to compute quantum corrections to
the \Kh\ potential at one and two loops. To obtain the functional
dependence of one and two loop corrections on chiral multiplets,
we expand the theory around some non--trivial background 
$\gf$ for the chiral multiplets, while assuming a trivial background
for the gauge sector. We do not need a non--trivial gauge background
because the appearance of the vector multiplet in the effective \Kh\
potential can be reconstructed by gauge invariance. The background
$\gf$ is assumed to fulfill the classical equations of motion. To
expand around this background we make the  replacement $\gf ~\ra~ \gf
+ \gF\,$, and declare that  the vector superfield $V$ and the chiral
superfield $\gF$ are quantum fields. Only the quantum fields appear in
the loops, while the background  fields encoded the external legs.

For generic non--renormalizable models also higher derivative
operators will be required as counter terms. Since we only focus on
the computation of the quantum effective \Kh\ potential not the full
effective quantum action, we will be ignoring all (super covariant)
derivatives ($D_\ga$, $\bD_\dga$ and $\der_m\,$) on the background
fields. It is irrelevant whether these derivatives arises from partial
integrations or the use of the background equations of motion. Let us
stress, that this does not mean that we treat $\gf$ as a strictly
constant background. Because if we would take $\gf$ constant, any
superspace integral over them will vanish, including the effective
\Kh\ potential which we seek to compute.  But it does mean that in 
our computation we are allowed to ignore the super derivative terms of
the background equations of motion 
\equ{
- \frac 14 \bD{}^2 K_a + W_a ~=~ 0~, 
\qquad 
- \frac 14 D{}^2 K^\ua + \bW^\ua ~=~ 0~, 
\labl{EoM} 
}
because they would give rise to higher derivative operators. 
(Here differentiations w.r.t.\ the background variables $\gf$ and
$\bgf$ are denoted as $K_a ~=~ K_{,a}$, $K^{\ua} ~=~ K_,{}^{\ua}\,$,
etc.)  Therefore, for our purposes we may use that $W_a ~=~ 0\,,$ even
though the field value of $\gf$ does not need to satisfy this
supersymmetric vacuum condition.

In the expansion around the background $\gf$ and in the computation of
the one and two loop effective \Kh\ potential we run into various
geometrical quantities, which we introduce here. The \Kh\ metric is
defined as $G^\ua{}_a ~=~ K^\ua{}_a ~=~ K_,{}^\ua{}_a\,;$ 
with $G\inv$ we indicate its inverse. Given the metric we can
construct other geometrical quantities like the connections 
\equ{
\gG_{ab}^{~c} ~=~ 
K_{ab}{}^\uc \,G\inv{}^c{}_{\uc}~, 
\qquad  
\bgG^{\ua\,\ub}_{~\,\, \uc} ~=~ 
K^{\ua\,\ub}{}_c \, G\inv{}^c{}_{\uc}~,
\labl{connections} 
}
and the curvature 
\equ{
R^\ua{}_a{}^\ub{}_b ~=~ 
K^{\ua\,\ub}{}_{ab} 
- K^{\ua\,\ub}{}_c \, G\inv{}^c{}_\uc \, K_{ab}{}^\uc~. 
\labl{curvature} 
} 
Using the connections we can define covariant derivatives 
$W_{;ab\dots}$ of the superpotential $W\,$. As explained around
\eqref{EoM} we may set $W_a ~=~ 0$ to find 
\equ{
W_{;ab} ~=~ W_{ab}~, 
\qquad 
W_{;abc} ~=~ W_{abc} - \gG_{ab}^{~d} \, W_{dc} 
 - \gG_{bc}^{~d} \, W_{da}  - \gG_{ca}^{~d} \, W_{db}~.  
\labl{covderW} 
}
This completes the classical geometrical description of the globally
supersymmetric linearly--gauged non--linear sigma model.

To define the perturbative quantum theory we perform the expansion in
quantum field $\gF$ and $V$ around the background $\gf$. 
The non gauge fixed action for the quantum fields $\gF$ and $V$ is
composed of the (classical) \Kh\ term 
\equ{
S_{K} ~=~ \frac 12 \,\int \d^8 z\, \Big\{ 
K \Big( \bgf + \bgF, \, e^{2V} (\gf + \gF)\Big) 
+ 
K \Big( (\bgf + \bgF) e^{2V}, \gf + \gF\Big) 
+ \gx \, \tr V 
\Big\}
~,
\labl{clKh} 
} 
with the superspace measure $\d^8 z ~=~ \d^4 x \,\d^4 \gth\,,$ a
Fayet--Iliopoulos parameters $\gx\,$, the superpotential
interactions    
\equ{
S_W ~=~ 
\int \d^6 z\, W(\gf + \gF) ~+~ \int\d^6 \bz\, \bW(\bgf+\bgF)~, 
\labl{clW}
} 
and finally the gauge kinetic part  
\equ{ 
S_G ~=~ \frac 14 \int\d^6 z\,\tr\,  f_{IJ}(\gf+ \gF) \, \cW^{I\,\ga} \cW^{J}{}_\ga ~+~ 
\frac 14 \int\d^6 z\,\tr\,  \bar f_{IJ} (\bgf+ \bgF) \, \bcW^{I}{}_\dga
\bcW^{J\,\dga}~, 
\labl{clG}
} 
with the chiral measure $\d^6z ~=~ \d^4 x\d^2\gth$ and is conjugate. 
The gauge kinetic function $f_{IJ}$ is symmetric in the gauge indices
$I$ and $J$. Its real part $h_{IJ} ~=~ \frac 12 ( f_{IJ} + \bar f_{IJ} )$ 
is equal to the gauge coupling $1/g^2\,$. A non--constant gauge kinetic
function appears for example in the effective models from string
theory, in particular when they rely on the Green--Schwarz
mechanism to cancel anomalies. The super gauge field strengths are
given by  
\equ{
\cW_\ga ~=~ -\frac 18\,{\bD{}^2} \Big( e^{-2V} D_\ga e^{2V}
\Big)~,
\qquad 
\bcW_\dga ~=~ -\frac 18\, {D^2}\Big( \bD_\dga e^{2V} e^{-2V}
\Big) ~, 
} 
with the notational convention that operators, like $D_\ga\,$, only act
on the first superfield on the right. We have given a representation of
the gauged \Kh\ potential action \eqref{clKh} which is manifestly real.

From these expression we determine the propagators of the various
superfields. Because of the super gauge invariance the kinetic
operator of the vector multiplet is not invertible. This requires 
gauge fixing and the introduction of the corresponding supersymmetric
Faddeev--Poppov ghosts $C,C', \bC, \bC'\,$, see e.g.\
\cite{Gates:1983nr,West:1990tg}.   
For the general supersymmetric theories under consideration in this
paper, two additional complications arise: Firstly, the background
$\gf$ spontaneously breaks some of the gauge symmetries, and therefore
leads to quadratic mixing of the vector multiplets $V$ with the
chiral multiplets $\gF$ and $\bgF\,$. To avoid this mixing the gauge
fixing functional $\gTh^I$ can be modified to  
\equ{
\gTh^I ~=~ -\frac {\sqrt 2}4 \bD^2\Big( 
V^I  
~+~ (h\inv)^{IJ} K^{\ua}{}_a (T_J \gf)^a \, \frac 1{\Box} \bgF_\ua
\Big)~. 
\labl{GFfunction}
}
This is very similar to the 't Hooft $R_\gx$ gauge fixing for
spontaneously broken gauge theories. The gauge fixing procedure is
then implemented in the standard way by the insertion of  
\equ{
\gD_{FP}\, \Big| \,\gd(\gTh^I - F^I) \, \Big|^2\, 
e^{i S_F}~,
\qquad 
S_F = \int d^8 z\, h_{IJ}\, \bF^I F^J
~, 
} 
in the path integral defining the quantum theory, where $\gD_{FP}$ is
the Faddeev--Poppov determinant and $F^I$ are arbitrary chiral
multiplets. The second complication is that  the Gaussian
integral over the gauge fixing chiral variables $F^I$ and $\bF^I$
involves the function $h_{IJ}$ which is in general a function of $\gf$
and its  conjugate. Therefore, we need to properly normalize this
Gaussian integral. This can be implemented by the introduction of the 
Nielsen--Kallosh ghosts $\gch^I\,$, which have the same quadratic
action as $F^I$, but they are anti--commuting.

The gauge fixing procedure together with the introduction of the
various ghosts leads to the addition of the following terms to the
classical action. First of all, the gauge fixing term is obtained by
performing the functional integral over $F^I\,$: 
\equ{
S_{GF} = - \int \d^8 z\, h_{IJ} \, \bgTh{}^I \gTh^J~, 
\labl{GFac}
} 
with $\gTh^I$ given in \eqref{GFfunction}. This seems to be worrying
because the presence of $1/\Box$ leads to non--local
interaction. However, as far as the definition of the perturbation
theory is concerned this non--locality leads to perfectly well defined
propagators. The Faddeev--Poppov ghost action reads 
\equ{
S_{FP} ~=~ \frac 1{\sqrt 2} \int \d^6 z \, C_I' \gd_C \gTh^I + 
\frac 1{\sqrt 2} \int \d^6 \bz \, \bC_I' \gd_C \bgTh^I~.
\labl{FPghosts}
} 
Here $\gd_C \gTh^I$ denotes the gauge transformation of the gauge
fixing functional to first order in the super gauge parameters 
\equ{
\gd_\gL \gTh^I ~=~ \sqrt{2}\, \frac {\bD^2}{-4} \Big\{
\bgL^I + [V, \gL - \bgL]^I 
- 2 (h\inv)^{IJ} (T_J \gf)^a K^\ua{}_a 
\frac 1{\Box} \big( \bgf\,\bgL  + \bgF \, \bgL\big)_\ua
\Big\} + \ldots~, 
\labl{VarGF} 
} 
but with the super gauge parameter $\gL^I$ replaced by the ghosts
$C^I\,$, etc. The dots denote higher order terms in $V$ which are not
relevant for our computations. To obtain this we have made use of the
super gauge transformations   
\equ{
\bgF ~\ra~ (\bgf + \bgF ) e^{-2\bgL} - \bgf~, 
\qquad 
e^{2V} ~\ra~ e^{2\bgL} \, e^{2V} \, e^{2\gL}~, 
\qquad 
\gF ~\ra~ e^{-2 \gL} (\gf + \gF) - \gf~, 
\labl{supergauge}
}
linearized in the (anti--)holomorphic super gauge parameters ($\bgL$)
$\gL\,$. Because the background generically spontaneously
breaks the gauge symmetry, the gauge transformations of the quantum
fields $\gF$ are non--linear.  Finally, the Nielsen--Kallosh ghost
action  
\equ{
S_{NK} ~=~ - \int \d^8 z\, h_{IJ}\, \bgch{}^I \gch^J
\labl{NKac}
} 
completes the full action of the quantum theory.

We now determine the propagators by reading off quadratic parts of the
quantum action. As the Nielsen--Kallosh ghost action \eqref{NKac} is 
quadratic by itself we do not need to discuss it again. The quadratic
actions of the vector and Faddeev--Poppov ghost superfields are given
by 
\equ{
\arry{l}{\dsp 
S_{V\phantom{P}}^2 ~=~ 
- \int\d^8 z\, V^I \, [h\, \Box - M_V^2]_{IJ} V^J~, 
\\[2ex] \dsp 
S_{FP}^2 ~=~ \phantom{-} \int\d^8 z\, \Big( 
C_I' \, \Big[ \Id - h\inv M_C^2 \, \frac 1\Box \Big]^I_{~J} \,\bC{}^J
+ 
\bC_{I}' \,  \Big[ \Id - h\inv {M_C^2}^T\, \frac 1\Box \Big]^I_{~J} \,C^J
\Big)~. 
} 
\labl{Ghostsqu} 
}
Here we have introduced the Hermitean mass matrices 
\equ{
(M_C^2)_{IJ} ~=~ 2\,  \bgf T_I G T_J \gf~, 
\qquad 
M_V^2 ~=~ \frac 12 \Big( M_C^2 + {M_C^2}^T \Big)~, 
\labl{VectorMM}
} 
for the ghost and vector multiplets. Because ${M_C^2}^T$ refers to the
transpose of $M_C^2\,$ (i.e.\ its vector indices $I,J$ are
interchanged), the matrix $M_V^2$ is symmetric. By
construction of the gauge fixing \eqref{GFfunction}
and \eqref{GFac} there is no mixing between the vector and the chiral
multiplet,  instead the quadratic part of the chiral multiplet action
has become more complicated  
\equ{
S_\gF^2 ~=~ \int \d^8 z\, \Big( 
\bgF_\ua \, \Big[ G - G M_G^2 \, \frac 1{\Box} \Big]^\ua_{~a} \, \gF^a
~+~ 
\frac 12\, \gF^a \, W_{ab} \frac{D^2}{-4\Box}\, \gF^b 
~+~ 
\frac 12\, \bgF_\ua \, \bW^{\ua\,\ub} \frac{\bD^2}{-4\Box}\, \bgF_\ub 
\Big)~, 
\labl{Chiralqu}
} 
where we introduce (partly for later reference) the superpotential,
Goldstone and total mass 
matrices 
\equ{
M_W^2 ~=~ G\inv \bW (G\inv)^T W~,
\quad 
(M_G^2)^a{}_b ~=~  
2\, (T_I \gf)^a\, (h\inv)^{IJ}\,  (\bgf T_J G)_b~, 
\quad 
M^2 ~=~ M_W^2 + M_G^2~, 
\labl{chMasses} 
}
respectively. We use the matrix notation $W$ to denote the second
derivative of the superpotential: $[W]_{ab} ~=~ W_{,ab}$ for
notational convenience.

\begin{figure}
\begin{center} 
\tabu{ccccccc}{
\raisebox{0ex}{\scalebox{0.3}{\mbox{\input{propbgFgF.pstex_t}}}}
&$\qquad$ & 
\raisebox{0ex}{\scalebox{0.3}{\mbox{\input{propgFgF.pstex_t}}}}
&$\qquad$ & 
\raisebox{0ex}{\scalebox{0.3}{\mbox{\input{propVV.pstex_t}}}}
&$\qquad$ & 
\raisebox{0ex}{\scalebox{0.3}{\mbox{\input{propbCC.pstex_t}}}}
}
\end{center} 
\captn{Here we depict the super propagators that will be employed to
  computed the quantum corrections to the \Kh\ potential. The first
  two diagrams correspond to the chiral propagators defined in
  \eqref{ChiralProp}: The first one represents $\gD_{\bgF\gF}$ and the
  second one $\gD_{\gF\gF}\,$. The latter two refer to the 
  vector and Faddeev--Poppov ghost propagators given in
  \eqref{GaugeProp}, respectively. }
\labl{fg:Props}
\end{figure}

From the quadratic expressions for the gauge and ghost multiplets it
is straightforward to determine the resulting superfield
propagators. Even though the quadratic action of the ghost multiplets 
\eqref{Ghostsqu} have non--local terms, the resulting propagators are
well--behaved 
\equ{
\gD_{VV} ~=~ [h \, \Box- M_V^2]\inv~, 
\qquad 
\gD_{C'\bC} ~=~ [ \Box- h\inv M_C^2]\inv~, 
\qquad 
\gD_{\bC' C}~=~ [ \Box- h\inv {M_C^2}^T]\inv~. 
\labl{GaugeProp}
} 
The non--locality of \eqref{Chiralqu} is also not reflected  in the
chiral multiplet propagators  
\equ{
\arry{c}{\dsp 
\gD_{\bgF\gF} 
~=~  [ \Box - M_G^2]\inv \, G\inv ~+~  
[ \Box - M_W^2]\inv\, G\inv ~-~  
G\inv \, \frac 1\Box 
~=~ [ \Box - M^2]\inv \, G\inv, 
\\[2ex] \dsp 
\gD_{\gF\gF} ~=~ 
G\inv \, [ \Box - M^2]\inv \,\bW (G\inv)^T 
~,\quad 
\gD_{\bgF\,\bgF} ~=~ 
(G\inv)^T W G\inv\,  [ \Box - M^2]\inv 
~.   
}
\labl{ChiralProp}
}
In supergraphs the propagators $\gD_{\gF\gF}$ and $\gD_{\bgF\,\bgF}$ are
multiplied by $-D^2/(4\Box)$ and $-\bD{}^2/(4\Box)\,$, respectively. 
(In figure \ref{fg:Props} we have collected our graphical
representation for these propagators.) Because of super gauge
invariance of the superpotential \eqref{GaugeInv}, the superpotential
matrix $W$ has zero modes  $T_I\gf\,$:  
\(
W_{ab} \, (T_I \gf)^b ~=~ 0\,,
\) 
up to derivative terms which we ignore in the computation of the
effective \Kh\ potential using the background equations of motion. 
By combining this observation with the definition of the masses in 
\eqref{chMasses} we arrive at the propagators \eqref{ChiralProp}. These 
zero modes  correspond to the Goldstone modes of the global symmetries 
generated by $T_I\,$. This implies that  
$M_W^2 \, M_G^2 ~=~ M_G^2 \, M_W^2 ~=~ 0\,$, 
and consequently that three propagators in the first equation of
\eqref{ChiralProp} can be combined into one, and that $M^2$ can be
replaced by $M_W^2$ in the last two equations of \eqref{ChiralProp}.

This can be interpreted as follows: The background defined by $\gf$
generically leads to spontaneous symmetry breaking and massive vector
multiplets. The corresponding massless Goldstone superfields are eaten
by these massive vector superfields in the unitary gauge. However, we
have chosen a different gauge, defined by \eqref{GFfunction} and
\eqref{GFac}, to ensure the absence of mixing between the vector and
the chiral multiplets. In this gauge a massive vector multiplet
consists of $V^I$ (with $I$ such that $T_I \gf \neq 0\,$), the
Goldstone mode chiral superfields (identified by $T_I \gf\,$), and the
massive Faddeev--Poppov ghosts $C_I, C_I'\,$. Moreover, in this gauge
the chiral Goldstone multiplets and the ghost multiplets have the same
mass eigenvalues, because for any integral power $p$ we have  
\equ{
\tr (M_G^2)^p ~=~ \Tr (h\inv M_C^{2})^p
~=~ \Tr (h\inv {M_C^2}^T)^p\,. 
\labl{relationtraces}
}
Moreover, the symmetric part of the ghost mass is equal to the vector
multiplet mass, see \eqref{VectorMM}. This leads to partial
cancellations of the contributions  of the Goldstone chiral multiplets
and the Faddeev--Poppov ghosts in the one loop computation of the \Kh\
potential, as we will see in the next section.


%% file: propbgFgF.pstex_t
\begin{picture}(0,0)%
\includegraphics{propbgFgF.pstex}%
\end{picture}%
\setlength{\unitlength}{4144sp}%
\begingroup\makeatletter\ifx\SetFigFont\undefined%
\gdef\SetFigFont#1#2#3#4#5{%
  \reset@font\fontsize{#1}{#2pt}%
  \fontfamily{#3}\fontseries{#4}\fontshape{#5}%
  \selectfont}%
\fi\endgroup%
\begin{picture}(1844,314)(2004,-3818)
\end{picture}

%% file: propgFgF.pstex_t
\begin{picture}(0,0)%
\includegraphics{propgFgF.pstex}%
\end{picture}%
\setlength{\unitlength}{4144sp}%
\begingroup\makeatletter\ifx\SetFigFont\undefined%
\gdef\SetFigFont#1#2#3#4#5{%
  \reset@font\fontsize{#1}{#2pt}%
  \fontfamily{#3}\fontseries{#4}\fontshape{#5}%
  \selectfont}%
\fi\endgroup%
\begin{picture}(1844,314)(2004,-3818)
\end{picture}

%% file: propVV.pstex_t
\begin{picture}(0,0)%
\includegraphics{propVV.pstex}%
\end{picture}%
\setlength{\unitlength}{4144sp}%
\begingroup\makeatletter\ifx\SetFigFont\undefined%
\gdef\SetFigFont#1#2#3#4#5{%
  \reset@font\fontsize{#1}{#2pt}%
  \fontfamily{#3}\fontseries{#4}\fontshape{#5}%
  \selectfont}%
\fi\endgroup%
\begin{picture}(1844,318)(1959,-3685)
\end{picture}

%% file: propbCC.pstex_t
\begin{picture}(0,0)%
\includegraphics{propbCC.pstex}%
\end{picture}%
\setlength{\unitlength}{4144sp}%
\begingroup\makeatletter\ifx\SetFigFont\undefined%
\gdef\SetFigFont#1#2#3#4#5{%
  \reset@font\fontsize{#1}{#2pt}%
  \fontfamily{#3}\fontseries{#4}\fontshape{#5}%
  \selectfont}%
\fi\endgroup%
\begin{picture}(1844,44)(2004,-3683)
\end{picture}

%% file: oneloop.tex
\newpage 
\section{One loop effective \Kh\ potential}
\labl{sc:oneloop}

The one loop calculation of the effective \Kh\ potential involves the
computation of one loop vacuum bubble graphs with multiple insertions
of two--point interaction terms, as indicated in figure \ref{fg:1Lbubble}. 
The definition of the two point interaction $\cM$ and the propagator
$\gD$ here is somewhat arbitrary. To see why this is actually
convenient, consider for example, a chiral multiplet with both a
mass term and a quadratic \Kh\ term,  which are both functions of the
background fields. To evaluate the corresponding Gaussian integral, we
can first take the mass as a two point interaction and compute in the
way explained below. The remaining Gaussian integral with a \Kh\
metric between the quantum chiral superfields is evaluated by taking
the identity matrix to define the propagator, and the \Kh\ metric
minus the identity as the two point interaction.  Hence, it is
precisely this freedom in the choice of the propagator and the two
point interaction that helps to make convenient choices to evaluate
the Gaussian path integrals effectively.

To evaluate these bubbles in general, we consider a generic vector of
commuting superfields $\gPs$ with quadratic action  
\equ{
S ~=~ \frac 12\,  \int\d^8 z\, \gPs^T  \, [ \gD\inv + \cM] \, \gPs~.
} 
The sum of the connected bubble graphs reads 
\equ{
i \gG ~=~ \frac 12  \sum_{n \geq 1} \frac{(-)^n}{n}  
\int (\d^8 z)_{\dots} \, \tr 
\Big( 
\cM_1 X_{\ubar{1}1} \gD_{\ubar{1}2} ~ 
\cM_2 X_{\ubar{2}2} \gD_{\ubar{2}3} ~ 
\ldots ~ \cM_n X_{\ubar{n}n} \gD_{\ubar{n}1}
\Big)~, 
\labl{BubbleSum} 
} 
where $1,2, \ldots, n$ and $\ubar{1},\ubar{2},\ldots,\ubar{n}$ are 
labels that denote the various superspace coordinates 
$z_1,z_2, \ldots, z_n$ and $z_\ubar{1},z_\ubar{2}, \ldots,z_\ubar{n}\,$.
The corresponding measure is represented by $(\d^8 z)_{\ldots}\,$. 
By functional differentiation w.r.t.\ sources $J$ that couple to
$\gPs\,$, we have obtained 
\equ{
X_{\ubar{1} 1} ~=~ \frac{\gd \, J_{\ubar{1}}}{\gd\, J_1} 
~=~ \frac{\gd \, J(z_{\ubar{1}})}{\gd\, J(z_1)} 
~=~ \gd_{\ubar{1}1}~. 
}
Here 
$\gd_{\ubar{1}1} ~=~ \gd^4_{\ubar{1}1} \gd^4(\gth_\ubar{1}-\gth_1)$
denotes the superspace delta function,  and 
$\gd_{\ubar{1}1}^4= \gd^4(x_\ubar{1}-x_1)$ the four dimensional 
spacetime delta function. (When the sources correspond to chiral
multiplets, we find some additional super covariant derivatives
hitting the superspace delta function.) When we consider
anti--commuting superfields, the Faddeev--Popov and Nielsen--Kallosh
ghosts, the expression in \eqref{BubbleSum} has one additional overall
minus sign.

\begin{figure}
\begin{center} 
\raisebox{0ex}{\scalebox{0.45}{\mbox{\input{bubbles.pstex_t}}}}
\end{center} 
\captn{At the one loop level the effective \Kh\ potential is obtained
  from summing an infinite set of one loop bubble diagrams. This
  corresponds to the computation of various functional determinants. 
}
\labl{fg:1Lbubble}
\end{figure}

We can apply this to the various quadratic terms derived in the previous
section. Using various superspace and supergraph techniques we finally
find that the full one loop \Kh\ potential is given by 
\equ{
i \gG_{1L} = \int (\d^4 x)_{12}\d^4 \gth\, \Big[ 
\Tr \ln h + \Tr \ln \Big( \Id - \frac{h\inv M_C^2}{\Box} \Big) 
- \tr \ln G 
- \frac 12\, \tr\ln \Big(\Id - \frac{M_W^2}{\Box} \Big)
\Big]_1 \gd_{12}^4 \, \frac 1{\Box_1} \gd_{12}^4~,
\labl{OneKh}
} 
in a coordinate representation. In accordance with a general
non--renormalization theorem of supersymmetry
\cite{Grisaru:1979wc,West:1990tg}, this expression is local in the
Grassmann variables. This is because of the superspace expression
becomes proportional to  
\(
A_2\, \gd_{12} \, [B \,D^2\bD{}^2]_2 \, \gd_{12} ~=~ 
16\, A_2\, \gd^4_{12} \, B_2\, \gd_{12}^4\, 
\gd^4(\gth_1-\gth_2)\,,
\)
where $A$ and $B$ are some operators which may be functions of
spacetime derivatives $\der_m\,$. The origins of the various terms in
\eqref{OneKh} are as follow:  
The first term is due to the Nielsen--Kallosh ghosts. The second term
is the combined effective action of the Faddeev--Poppov ghosts and the
Goldstone chiral multiplets, using that their mass eigenvalues in the
gauge \eqref{GFfunction} are all equal, see \eqref{relationtraces}. 
(The bubble graphs with vector multiplets vanish identically in this
gauge, because no $D$, $\bD$ ever hit any superspace delta functions.)
The last two terms are due bubbles that contain chiral multiplets.

As it stands this expression \eqref{OneKh} is ill--defined and
requires regularization. Mainly because computational convenience at the
two loop level, we have chosen to use dimensional reduction 
\cite{Siegel:1979wq,Capper:1979ns}. In appendix \ref{sc:1Lintegrals}
we have collected the one loop integrals calculated in this scheme. A
peculiar consequence of the \DRbar\ scheme is, that both the first and
the third term are absent, as follows from \eqref{J1Lint}. Using
\eqref{Log1Lint} and \eqref{Log1LintRes}, and dropping the $1/\ge$
poles, we find that the remaining terms result in the effective one
loop \Kh\ potential 
\equ{
K_{1L} ~=~ 
- \frac 1{16\gp^2} \, 
\Tr\, h\inv M_C^2 \Big( 2 - \ln \frac {h\inv M_C^2}{\bgm^2} \Big)
~+~ \frac 1{32\gp^2}\, 
\tr\, M_W^2 \Big( 2 - \ln \frac {M_W^2}{\bgm^2} \Big)
~, 
\labl{1LKh}
} 
where $\bgm^2 ~=~ 4\gp e^{-\gg} \gm^2$ defines the \MSbar\
renormalization scale, and the mass matrices $M_C^2$ and $M_W^2$ are
defined in \eqref{VectorMM} and \eqref{chMasses}, respectively. 
To use the expressions from appendix \ref{sc:1Lintegrals} for
matrices, we can first diagonalize  the mass matrices, and apply these
expressions for the eigenvalues, and after that transform back to the
original basis.

\begin{figure}
\begin{center} 
\raisebox{0ex}{\scalebox{0.5}{\mbox{\input{1Lself.pstex_t}}}}
\end{center} 
\captn{This one loop self energy supergraph is computed as a cross
  check for the one loop \Kh\ potential determined in this section.
}
\labl{fg:1LSelfEnergy}
\end{figure}

To concluded the discussion concerning the one loop effective \Kh\
potential, we would like to mention an important consistency check of
our computation and compare our results to the existing literature.
The effective \Kh\ potential can be used to determine the wave function
renormalization at one loop by taking the second mixed derivative of
it. This wave function renormalization can also be computed directly
from the supergraph given in figure \ref{fg:1LSelfEnergy}. For this
cross check it is sufficient to consider the renormalizable
Wess--Zumino model (see section \ref{sc:nonWZ} for details). The self
energy supergraph gives the wave function renormalization 
\equ{
\gS_{\text{self energy}} ~=~ 
- \frac {|\gl|^2} {32 \gp^2}\, \ln \frac {|m + \gl\, \gf|^2}{\bgm^2}~, 
} 
which agrees with our one loop effective \Kh\ potential result: 
\equ{
\gS_{\text{eff.\ \Kh\ pot.}} ~=~
\frac{\der^2 \, K_{1L}}{\der \gf \, \der \bgf} ~=~ 
- \frac {|\gl|^2} {32 \gp^2}\, \ln \frac {|m + \gl\, \gf|^2}{\bgm^2}~, 
}
see \eqref{KHrenWZ}. As long as no non--Abelian gauge interaction are
taken into account, our one loop results are consistent with
some existing literature concerning the computation of the effective
\Kh\ potential to that order
\cite{Grisaru:1996ve,deWit:1996kc,Pickering:1996gt,Buchbinder:1998qv,Buchbinder:1999jw,Brignole:2000kg}.
In the non--Abelian case the mass matrices $M_C^2$ and $M_V^2$ are not
equal anymore, and our results slightly deviate from these reference. 
This might be an artifact of the use of different gauge fixing
procedures.


%% file: bubbles.pstex_t
\begin{picture}(0,0)%
\includegraphics{bubbles.pstex}%
\end{picture}%
\setlength{\unitlength}{4144sp}%
\begingroup\makeatletter\ifx\SetFigFont\undefined%
\gdef\SetFigFont#1#2#3#4#5{%
  \reset@font\fontsize{#1}{#2pt}%
  \fontfamily{#3}\fontseries{#4}\fontshape{#5}%
  \selectfont}%
\fi\endgroup%
\begin{picture}(4291,928)(4216,-5115)
\end{picture}

%% file: 1Lself.pstex_t
\begin{picture}(0,0)%
\includegraphics{1Lself.pstex}%
\end{picture}%
\setlength{\unitlength}{4144sp}%
\begingroup\makeatletter\ifx\SetFigFont\undefined%
\gdef\SetFigFont#1#2#3#4#5{%
  \reset@font\fontsize{#1}{#2pt}%
  \fontfamily{#3}\fontseries{#4}\fontshape{#5}%
  \selectfont}%
\fi\endgroup%
\begin{picture}(1664,772)(2679,-4497)
\end{picture}

%% file: twoloop.tex
\section{Two loop effective \Kh\ potential}
\labl{sc:twoloop}

At the two loop level there are three different topologies of the
supergraphs that may contribute to the \Kh\ potential. They are
depicted in figure \ref{fg:2Lgraphs}. The third topology is
one--particle--reducible and can be ignored when computing 
the effective \Kh\ potential as we will argue in subsection
\ref{sc:doublegraphs}. In the following two subsections we first
reduce the supergraphs to scalar momentum integrals. Then using the
evaluation of these momentum integrals in appendix 
\ref{sc:2Lintegrals}, we can give the renormalized expressions for
these scalar integrals in the \DRbar\ scheme. We collect our final
results in subsection \ref{sc:summary2L}.

We obtain the vertices, from which the ``8'' and ``$\ominus$''
graphs can be derived, by expanding the various actions given in
section \ref{sc:setup} up to fourth order in the quantum fields. The
four point vertices are relevant for the ``8'' supergraphs, while
``$\ominus$'' supergraphs can be build from two three--point
interactions.  
We list only the relevant vertices for diagrams that do not vanish 
automatically. When performing these expansions in the quantum fields,
it is important to take the consequences of gauge invariance
\eqref{GaugeInv} of the \Kh\ potential $K$ and the superpotential $W$
into account.  Because none of the propagators \eqref{GaugeProp} and
\eqref{ChiralProp} mix with each other, we can exclude many diagrams,
which would be present in many other gauges. In addition we have the
constraint, that supergraphs are only non--vanishing at two loop if
they involve an equal number of super 
covariant derivatives $D$ and $\bD$, and of each of them at least
four. Chiral and ghost multiplets will introduce a combination of
$D^2$ and $\bD{}^2$ into the diagram, while the vector multiplets do
not. Therefore, the number of $D^2$ and $\bD{}^2$ of graphs with 
vector superfields will in general be lower than the ones involving
only chiral multiplets, and hence such graphs are more likely to
vanish. As we will see in the subsequent subsections such
considerations reduce the number of relevant diagrams, to a manageable
number for manual computations.

\begin{figure}
\begin{center} 
\tabu{ccccc}{
\raisebox{.5ex}{\scalebox{0.5}{\mbox{\input{sc8graph.pstex_t}}}}
&\qquad \qquad &
\raisebox{0ex}{\scalebox{0.3}{\mbox{\input{sc0-graph.pstex_t}}}}
&\qquad\qquad & 
\raisebox{0ex}{\scalebox{0.5}{\mbox{\input{scdoublegraph.pstex_t}}}}
\\[2ex]
a&& b &&c
}
\end{center} 
\captn{There are three basic vacuum graphs at the two loop
  level. They have the topologies of an ``8'' (figure a) and
  ``$\ominus$'' (figure b), a ``double tadpole'' (figure c),
  respectively.  
}
\labl{fg:2Lgraphs}
\end{figure}

\subsection{``Double tadpole'' supergraphs}
\labl{sc:doublegraphs}

Most computations of the effective (\Kh) potential are restricted 
to only those connected graphs that are one--particle--irreducible. 
The argument for this restriction is that all one--particle--reducible
connected graphs contain one or more tadpole subgraphs, which are
generically absent by symmetry arguments. (For example, a $\gf^4$
theory has the symmetry $\gf ~\ra~ -\gf$ which forbids tadpoles to
arise.) Because we are dealing with rather generic supersymmetric
models in arbitrary backgrounds, we reconsider the issue of
one--particle--reducible graphs.

For the computation of the effective \Kh\ potential at two
loops there is only a single one--particle--reducible topology, which
has been depicted in figure \ref{fg:2Lgraphs}.c. The connecting line
can represent either a chiral or a vector multiplet propagators, while
the loop lines may also refer to Faddeev--Poppov ghost multiplets.
We can divide these diagrams into two classes depending on whether the
connecting line is a chiral or a vector multiplet. In the case that
the connecting line is a chiral superfield, one can show by some
partial integrations of $D^2$ or $\bD{}^2$ within the diagram and
using relations like $D^2\bD{}^2 D^2 ~=~ 16 \Box\, D^2\,$, that
these diagrams contain too little $D^2$ or $\bD{}^2\,$, and therefore
vanish.

This leaves us with double tadpole graphs (figure
\ref{fg:2Lgraphs}.c) with a vector multiplet as a connecting line. 
Because a vector multiplet is not chiral, no $D^2$ or $\bD{}^2$ appear
on the connecting line. This implies that these double tadpole
graphs are non--vanishing iff the sum of Fayet--Illiopoulos tadpole
graphs a vector superfield is non--zero. The conditions for this have
been well--studied in the literature (see e.g.\ ref.\
\cite{Fischler:1981zk} and Weinberg's third volume
\cite{Weinberg:2000cr}).  Let us briefly review the arguments which
are applicable in our case: If the vector multiplet is non--Abelian
no tadpole is possible because the tadpole graph is never gauge
invariant because of higher order (vector multiplet $V$ dependent) Lie
derivatives acting on the chiral and anti--chiral gauge parameters
$\gL$ and $\bgL$. For a $\U{1}$ vector superfield $V$ a tadpole 
(see \eqref{clKh}) is possible provided that the Fayet--Iliopoulos
parameter $\gx$ is not a function of any (background) superfields. 
(Otherwise, again, the tadpole is not gauge invariant.) In particular,
the induced $\gx$ at the one loop level is a constant, which is
proportional to the sum of charges of all massless charged chiral
superfields times the integral  
\(
\int \d^4 p/p^2\,. 
\)
Since in this work we use dimensional reduction throughout, this
integral vanishes. Therefore we conclude that the double tadpole
graphs do not give any contribution to the effective \Kh\ potential at
two loops in the \DRbar\ scheme.

\subsection{Supergraphs of the ``8'' topology}
\labl{sc:8graphs}

\begin{figure}
\begin{center} 
\tabu{c}{
\raisebox{0ex}{\scalebox{0.5}{\mbox{\input{8supergraph.pstex_t}}}}
}
\end{center} 
\captn{This is the only not necessarily vanishing supergraph of the
  ``8'' topology. 
}
\labl{fg:8supergraph}
\end{figure}

The arguments presented in the first part of this section imply that
there is in fact only one ``$8$"  supergraph (see figure
\ref{fg:8supergraph}) that results from the 
vertex  
\equ{
\gD S^4 ~\supset~ \int \d^8 z\, 
\frac 14\, K_{ab}{}^{\ua\,\ub}\, \gF^a \gF^b\, \bgF_\ua \bgF_\ub~, \ 
\labl{4Point} 
}
which is obtained from the \Kh\ potential \eqref{clKh}. Using standard
supergraphs techniques we find that the supergraph, figure
\ref{fg:8supergraph}, becomes the following scalar integral  
\equ{
i \gG_{2L}^{\text{``8''}} ~=~ 
- \frac i2 \int (\d^4 x)_{123} \d^4\gth\, K_{1\, ab}{}^{\ua\,\ub} \, 
\gd_{21}^4 (\gD_{\gF\bgF})^a_{2\,\ua} \gd_{21}^4 \, 
\gd_{31}^4 (\gD_{\gF\bgF})^b_{3\,\ub} \gd_{31}^4~,
\labl{8supergraph}
}
with the chiral superfield propagator given in \eqref{ChiralProp}. The
subscript '1' on the fourth mixed derivative of the \Kh\ 
potential emphasizes that it is evaluated in coordinate system '1'. 
For the other subscripts we refer back to section \ref{sc:oneloop}.

By doing a Fourier transform to momentum space we see that also the
corresponding scalar integral has the ``8'' topology. Because we need
to do a double Wick rotation to Euclidean space, the expression
changes sign. The resulting scalar integral $J$ defined in \eqref{sc8}
is evaluated in appendix \ref{sc:2Lintegrals}. However, here the
masses in the propagators are 
matrices rather than ordinary numbers. This does not pose a problem
because we can simply reinterpret the formula for matrix valued
masses. (This can be confirmed by performing a diagonalization of the
masses after which the results of the appendix \ref{sc:2Lintegrals}
can be readily used.) Concretely, the ``8'' supergraph can be
compactly expressed as   
\equ{
i \gG_{2L}^{\text{``8''}} ~=~  \frac i2\, 
\int \d^8 z\, K^{\ua\,\ub}{}_{ab}\, 
\bJten(M^2, M^2)~,
\labl{8result}
} 
where the mass matrix $M^2$ is given by \eqref{chMasses} and 
$\bJten$ is defined in \eqref{bJten}. 
Notice that, this expression is not covariant. This signals that this
result is not complete. In the next section we see that covariance is
restored by contributions coming from ``$\ominus$'' supergraphs.

\subsection{Supergraphs of the ``$\boldsymbol{\ominus}$'' topology} 
\labl{sc:ominusgraphs} 

\newcommand{\ominusgraphs}{
\begin{figure}
\begin{center} 
\tabu{cc cc cc cc cc cc cc cc cc c}{
\raisebox{0ex}{\scalebox{0.3}{\mbox{\input{0-superA.pstex_t}}}}
& 
\raisebox{0ex}{\scalebox{0.3}{\mbox{\input{0-superB.pstex_t}}}}
& 
\raisebox{0ex}{\scalebox{0.3}{\mbox{\input{0-superC.pstex_t}}}}
& 
\raisebox{0ex}{\scalebox{0.3}{\mbox{\input{0-superD.pstex_t}}}}
&
\raisebox{0ex}{\scalebox{0.3}{\mbox{\input{0-superE.pstex_t}}}}
&
\raisebox{0ex}{\scalebox{0.3}{\mbox{\input{0-superF.pstex_t}}}}
& 
\raisebox{0ex}{\scalebox{0.3}{\mbox{\input{0-superG.pstex_t}}}}
& 
\raisebox{0ex}{\scalebox{0.3}{\mbox{\input{0-superH.pstex_t}}}}
& 
\raisebox{0ex}{\scalebox{0.3}{\mbox{\input{0-superI.pstex_t}}}}
& 
\raisebox{0ex}{\scalebox{0.3}{\mbox{\input{0-superJ.pstex_t}}}}
\\[2ex]
A & B & C & D & E & F & G & H & I & J
}
\end{center} 
\captn{
These are all the non--vanishing supergraphs of the ``$\ominus$''
topology that can be obtained from the interaction \eqref{3Point}. The
arrows indicate whether chiral or anti--chiral legs are attached a 
vertex. If there are only chiral lines at a vertex and all their arrows
are all either pointing inwards or outward, the vertex is a
superpotential interaction.  A thick square vertex refers to
interactions that arise from the gauge kinetic functions. All other 
interactions involve the \Kh\ potential. 
}
\labl{fg:0-supergraphs}
\end{figure}
}

\begin{figure}
\begin{center} 
\tabu{cc cc cc cc cc cc cc cc cc c}{
\raisebox{0ex}{\scalebox{0.3}{\mbox{\input{0-superA.pstex_t}}}}
& 
\raisebox{0ex}{\scalebox{0.3}{\mbox{\input{0-superB.pstex_t}}}}
& 
\raisebox{0ex}{\scalebox{0.3}{\mbox{\input{0-superC.pstex_t}}}}
& 
\raisebox{0ex}{\scalebox{0.3}{\mbox{\input{0-superD.pstex_t}}}}
&
\raisebox{0ex}{\scalebox{0.3}{\mbox{\input{0-superE.pstex_t}}}}
&
\raisebox{0ex}{\scalebox{0.3}{\mbox{\input{0-superF.pstex_t}}}}
& 
\raisebox{0ex}{\scalebox{0.3}{\mbox{\input{0-superG.pstex_t}}}}
& 
\raisebox{0ex}{\scalebox{0.3}{\mbox{\input{0-superH.pstex_t}}}}
& 
\raisebox{0ex}{\scalebox{0.3}{\mbox{\input{0-superI.pstex_t}}}}
& 
\raisebox{0ex}{\scalebox{0.3}{\mbox{\input{0-superJ.pstex_t}}}}
\\[2ex]
A & B & C & D & E & F & G & H & I & J
}
\end{center} 
\captn{
These are all the non--vanishing supergraphs of the ``$\ominus$''
topology that can be obtained from the interaction \eqref{3Point}. The
arrows indicate whether chiral or anti--chiral legs are attached a 
vertex. If there are only chiral lines at a vertex and all their arrows
are all either pointing inwards or outward, the vertex is a
superpotential interaction.  A thick square vertex refers to
interactions that arise from the gauge kinetic functions. All other 
interactions involve the \Kh\ potential. 
}
\labl{fg:0-supergraphs}
\end{figure}

Next we consider the more complicated diagrams of the ``$\ominus$''
topology. We first list the  three point vertices that can give rise
to non--vanishing supergraphs of this topology. From the \Kh\ potential
\eqref{clKh}, superpotential \eqref{clW}, the gauge kinetic terms
\eqref{clG} and the Faddeev--Poppov ghosts  \eqref{FPghosts} and
\eqref{VarGF} we obtain  
\equa{
\gD S_K^3 ~\supset& \,\dsp  \int \d^8 z\, 
\Big\{
K^\ua{}_{a} \, (V\gF)^a  \,(  \bgF + 2 \bgf V)_\ua
+ K_{ab}{}^\ua \, \gF^b  
\Big(
(\bgf V) _\ua (\gF + 2 V \gf)^a  + 
\bgF_\ua (V \gf + \frac 12 \gF) ^a  
\Big) 
\Big\}+ \text{h.c.} ~, 
\non \\[2ex] 
\gD S_G^3 ~\supset &\, \dsp 
-\, \int \d^8 z \,
\Big\{
\frac 18\, 
f_{IJ}\, \bD^2 D^\ga V^I [D_\ga V, V]^J 
~+~ \frac {1}{16}\, 
f_{IJ\,a} \, \gF^a D^\ga V^I \bD{}^2 D_\ga V^J 
\Big\} 
+ \text{h.c.}~, 
\labl{3Point} \\[2ex]
\gD S_W^3 ~\supset &\, \dsp 
\int \d^8 z \, \frac 1{3!}\, W_{abc}\,
 \gF^a \gF^b \frac{D^2}{-4\Box} \gF^c
+ \text{h.c.}~, 
\qquad 
\gD S_{FP}^3 ~\supset \, \dsp 
\int \d^8 z \, 
(C'+\bC')_I [V, C- \bC]^I 
~, \non 
}
respectively. The form of the triple interaction involving the \Kh\
potential has been obtained after extensive use of the consequence of
gauge invariance \eqref{GaugeInv}. The non--vanishing supergraphs are
collected in figure \ref{fg:0-supergraphs}.

As we did for the ``8'' supergraph in subsection \ref{sc:8graphs},
we will only give the expressions for the resulting scalar integrals
of the ``$\ominus$'' supergraphs. However, to illustrate how such two
loop supergraphs can be evaluated in general, we discuss the
supergraph expression of diagram A of figure \ref{fg:0-supergraphs} in
some detail. Moreover, this computation shows explicitly how the
non--covariance of the result \eqref{8result} of the ``8'' graph is
resolved. The expression for diagram A is given by 
\equ{
i \gG_{2L}^{\text{A}} ~=~ 
\frac i 2 \int(\d^8 z)_{12}\, 
K_{1\,ab}{}^\uc K^{\ua\,\ub}_{2~~c}\, 
\Big[ 
(\gD_{\gF\bgF})^a_{~\ua} \frac {\bD{}^2 D^2}{16}
\Big]_1 \gd_{12}\, 
\Big[ 
(\gD_{\gF\bgF})^b_{~\ub} \frac {\bD{}^2 D^2}{16}
\Big]_1 \gd_{12}\, 
\Big[ 
(\gD_{\gF\bgF})^c_{~\uc} \frac {D^2 \bD{}^2}{16}
\Big]_1 \gd_{12}~. 
\non 
}
To reduce this integral to a scalar integral we partially integrate
the $\bD{}^2 D^2$ that acts on the first $\gd_{12}\,$. Since throughout
this work we are only interested in the effective \Kh\ potential, we
systematically ignore all (super covariant) derivatives on background
superfield quantities. This implies, that $\bD{}^2 D^2$ act (in
opposite order) on the final $\gd_{12}$ to give: 
\(
D^2 \bD{}^2 \, D^2 \bD{}^2 \gd_{12} ~=~ 
16 \Box\, D^2 \bD{}^2 \gd_{12}\,, 
\) 
and hence this supergraph reduces to the scalar integral 
\equ{
i \gG_{2L}^{\text{A}} ~=~ 
\frac i 2 \int(\d^4 x)_{12}\d^4 \gth\,  
K_{ab}{}^\uc K^{\ua\,\ub}{}_{c}\, 
(\gD_{\gF\bgF})^a_{~\ua} \gd_{12}^4\, 
(\gD_{\gF\bgF})^b_{~\ub} \gd_{12}^4\, 
(\gD_{\gF\bgF}\Box)^c_{~\uc}  \gd_{12}^4~. 
\labl{Apreres}
} 
Since we are computing the effective potential  it is irrelevant
whether  $K_{ab}{}^\uc$ and the mass matrices inside the propagators
are functions of coordinate system '$1$' or '$2$', hence for
notational simplicity we have dropped the coordinate system labels. 
After the double Wick rotation, transforming to momentum space, and
computing the regularized Euclidean integrals, we obtain 
\equa{\dsp 
i \gG_{2L}^{\text{A}} ~= &\dsp \, 
- \frac i2 \int \d^8 z\, 
\Big\{ 
K^{\ua\,\ub}{}_c \, G\inv{}^c{}_{\uc}\, K_{ab}{}^{\uc} 
\, \bJten(M^2, M^2) 
- \bgG^{\ua\,\ub}_{~\, \ud} \, \bW^{\ud\,\uc}\, \gG_{ab}^{~d}\, W_{dc} 
\, \bIten(M^2,M^2,M^2) \Big\} 
\non \\[2ex] &\dsp \, 
+ i \int \d^8 z\, K^{\ua\,\ub}{}_{c} \, (T_I \gf)^{c} \,
K_{ab}{}^{\uc}\, (\bgf T_J)_{\uc} \, 
\bItenG(M^2,M^2,M_V^2)~. 
\labl{Ares} 
}
Here we have used \eqref{scOminusp^2}, and the notation defined in 
\eqref{bJten} and \eqref{bIten}. In the last term we have used
that the Killing vectors $T_I \gf$ are perpendicular to the
superpotential mass $M_W^2\,$, and therefore we may replace the full
mass matrix $M^2$ by Goldstone boson masses $M_G^2$. Moreover, by
using the relation between the Goldstone boson and the Faddeev--Poppov
ghost mass matrices, we can pull out the Killing vectors 
$(T_I \gf)^{c}$ and  $(\bgf T_J)_{\uc}\,$, and replace $M_G^2$ by
$M_C^2\,$.  Finally, because the last term is symmetric in the indices
$I, J\,$, we can replace $M_C^2$ by the vector multiplet mass matrix
$M_V^2\,.$  The derivation of \eqref{Ares} from \eqref{Apreres} 
illustrates, that even though diagram A has the ``$\ominus$'' topology
as a supergraph, it turns into a sum of ``$8$'' and ``$\ominus$''
scalar graphs. We see that the first term combines with the
contribution \eqref{8result}  of the ``$8$'' diagram to form the
curvature tensor \eqref{curvature}, and therefore leads to a covariant
expression as promised. The other parts of diagram A combine with the
other ``$\ominus$'' diagrams to give rise to covariant expressions as
well, as we will demonstrate below.

For the other supergraphs in figure \ref{fg:0-supergraphs} we only
give their reduction to scalar graphs and their final compact
expressions using the notation $\bI$ and $\bJ\,$, as they can be
computed using similar methods as diagram A above. All other
supergraphs reduce to scalar graphs of the  ``$\ominus$'' topology,
expect diagram H, which like diagram A, turns into both ``8'' and
``$\ominus$'' scalar graphs. 
To write down the results in a compact form, we employ the short hand 
notations 
\equ{
\arry{c}{
(G T_I \gf)^\ua{}_{;a} ~=~ 
K^\ua{}_b (T_I)^b{}_a + K^\ua{}_{ab} (T_I \gf)^b 
~=~
K^\ub{}_a (T_I)^\ua{}_\ub + K^{\ua\,\ub}_a (\bgf T_I)_\ub
~=~  (\bgf T_I G)_a{}^{;\ua}~, 
\\[2ex] 
(M_C^2)_{IJ}{}^{;\ua} ~=~ 
2(T_I G T_J \gf)^\ua + 2K^{\ua\,\ub}{}_{b} (\bgf T_I)_\ub (T_J \gf)^b~, 
\\[2ex]
(M_C^2)_{IJ}{}_{;a} ~=~ 
2(\bgf T_I G T_J)_a + 2K^{\ub}{}_{ab} (\bgf T_I )_\ub (T_J \gf)^b~. 
}
\labl{shortrelations}
}
The first equation in \eqref{shortrelations} is another consequence of
gauge invariance \eqref{GaugeInv} of the \Kh\ potential. Their
expressions for the other supergraphs of figure \ref{fg:0-supergraphs}
as scalar integrals are given by 
\equa{
i \gG_{2L}^{B} ~= & ~
i \int (\d^4 x)_{12}\d^4\gth\, 
K_{ab}{}^\uc \, K^{\ua\,\ub}{}_{c}\, 
(\gD_{\gF\bgF})^a{}_{\ua} \gd_{12}^4\, 
(\gD_{\gF\gF})^{bc}  \gd_{12}^4\, 
(\gD_{\bgF\,\bgF})_{\ub\,\uc}  \gd_{12}^4~, 
\non \\[2ex]
i \gG_{2L}^{C} ~= & ~
-\frac i2 \int (\d^4 x)_{12}\d^4\gth\, 
K_{ab}{}^\uc \, \bW^{\ua\,\ub\,\ud} \, 
(\gD_{\gF\bgF})^a{}_{\ua} \gd_{12}^4\, 
(\gD_{\gF\bgF})^{b}{}_{\ub}  \gd_{12}^4\, 
(\gD_{\bgF\,\bgF})_{\uc\,\ud}  \gd_{12}^4~, 
\non \\[2ex]
i \gG_{2L}^{D} ~= & ~
\frac i6 \int (\d^4 x)_{12}\d^4\gth\, 
W_{abc} \, \bW^{\ua\,\ub\,\uc} \, 
(\gD_{\gF\bgF})^a{}_{\ua} \gd_{12}^4\, 
(\gD_{\gF\bgF})^{b}{}_{\ub}  \gd_{12}^4\, 
(\gD_{\gF\bgF})^c{}_{\uc}  \gd_{12}^4~, 
\non \\[2ex]
i \gG_{2L}^{E} ~= & ~
- i \int (\d^4 x)_{12}\d^4\gth\, 
(G T_I \gf)^\ua{}_{;a} \, (\bgf T_J G)_b{}^{;\ub}\, 
(\gD_V)^{IJ} \gd_{12}\, 
(\gD_{\gF\bgF})^b{}_{\ua} \gd_{12}^4\, 
(\gD_{\gF\bgF})^{a}{}_{\ub}  \gd_{12}^4~,  
\non \\[2ex]
i \gG_{2L}^{F} ~= & ~
- i \int (\d^4 x)_{12}\d^4\gth\, 
K_{ab}{}^\uc (\bgf T_I)_\uc\, 
K^{\ua\,\ub}{}_c (T_I \gf)^c\, 
(\gD_V)^{IJ} \gd_{12}\, 
(\gD_{\gF\bgF})^a{}_{\ua} \gd_{12}^4\, 
(\gD_{\gF\bgF})^{b}{}_{\ub}  \gd_{12}^4~, 
\labl{0-graphs}  \\[2ex]
i \gG_{2L}^{G} ~= & ~
- \frac i4 \int (\d^4 x)_{12}\d^4\gth\, 
c^I{}_{JK} \, c^L{}_{MN}\, 
(\gD_V)^{MJ} \gd_{12}^4 \Big( 
(\gD_{C})^{K}{}_L \gd_{12}^4\, 
(\gD_{\bC})^{N}{}_I \gd_{12}^4 + 
\non \\ &
\qquad \qquad + 
(\gD_{\bC})^{K}{}_L \gd_{12}^4\, 
(\gD_{C})^{N}{}_I \gd_{12}^4 
- 
(\gD_{C})^{K}{}_L \gd_{12}^4\, 
(\gD_{C})^{N}{}_I \gd_{12}^4 
-
(\gD_{\bC})^{K}{}_L \gd_{12}^4\, 
(\gD_{\bC})^{N}{}_I \gd_{12}^4 
\Big)~, 
\non \\[2ex]
i \gG_{2L}^{H} ~= & ~
\frac i4 \int (\d^4 x)_{12}\d^4\gth\, 
f_{IJ\, a} \,f_{KL}{}^\ua \, 
(\gD_{V} \der_m)_1^{IK} \gd_{12}^4\, 
(\gD_{V} \der^m)_1^{JL} \gd_{12}^4 \, 
(\gD_{\gF\bgF})_1^a{}_{\ua} \gd_{12}^4~, 
\non \\[2ex]
i \gG_{2L}^{I} ~= & ~
\frac i2 \int (\d^4 x)_{12}\d^4\gth\, 
f_{IJ\, a} \, (M_C^2)_{KL}{}^{;\ua} \, 
(\gD_{V})_1^{IK} \gd_{12}^4\, 
(\gD_{V})_1^{JL} \gd_{12}^4 \, 
(\gD_{\gF\bgF})_1^a{}_{\ua} \gd_{12}^4~. 
\non  \\[2ex]
i \gG_{2L}^{J} ~= & ~
i \int (\d^4 x)_{12}\d^4\gth\, 
h_{LP}c^{P}{}_{IN} \, h_{JQ}c^{Q}{}_{KM}\, 
(\gD_{V})_1^{IJ} \gd_{12}^4\, 
(\gD_{V})_1^{KL} \gd_{12}^4 \, 
(\gD_{V})_1^{MN} \gd_{12}^4~. 
\non 
}
Notice that there are also the Hermitian conjugate diagrams of
diagrams C and I.

These expressions can be evaluated further using the same Fourier
transforms and Wick rotations as employed for diagram A. The diagrams
B, C, $\overline{\text{C}}$ and $D$ are all proportional 
the same integral, hence we can write 
\equa{
i \gG_{2L}^{\text{B}} + 
i \gG_{2L}^{\text{C}} + 
i \gG_{2L}^{\overline{\text{C}}} + 
i \gG_{2L}^{\text{D}} ~= ~
\frac i 6 \int \d^8 z\, 
\Big( & 
\bW^{\ua\,\ub\,\uc} \, W_{abc} 
-3 \, \bW^{\ua\,\ub\,\uc} \, \gG_{ab}^{~d}\, W_{dc} 
- 3 \, \bgG^{\ua\,\ub}_{~\, \ud} \, \bW^{\ud\,\uc}\, W_{abc} + 
\non\\[2ex] \dsp 
& + 6 \, 
\bgG^{\ua\,\ub}_{~\, \ud} \, \bW^{\ud\,\uc}\, \gG_{ab}^{~d}\, W_{dc}
\Big) 
\bIten(M^2,M^2,M^2)~. 
\labl{BCDres}
}
Observe that this combines with the parts of \eqref{Ares} that
proportional to the double derivatives of the superpotential to form
the square of the triple covariant derivative of the superpotential 
\eqref{covderW}.  From diagram E we get the contribution:  
\equ{
i \gG^{\text{E}}_{2L} ~=~  
- i \int \d^8 z\,  
(G T_I \gf)^\ua{}_{;b} \, 
(\bgf T_J G)_a{}^{;\ub} \, 
\bItenG(M^2,M^2,M_V^2)~. 
} 
It is not difficult to see that diagram F, given by 
\equ{
i \gG^{\text{F}}_{2L} ~=~ 
- i \int \d^8 z\, K^{\ua\,\ub}{}_{c} \, (T_I\gf)^{c} \, 
K_{ab}{}^{\uc}\, (\bgf T_J)_{\uc} \,
\bItenG(M^2,M^2,M_V^2)~,  
} 
is precisely opposite to the last contribution of diagram A in
\eqref{Ares}, so that these contributions precisely cancel each
other and covariance of the combined expression has become manifest. 
The effect of the ghosts results in diagram G 
\equ{
i \gG^{\text{G}}_{2L} = 
 \frac i2 \int \d^8 z\, h_{LP}c^P{}_{IN}\, h_{JQ}c^Q{}_{KM}\, 
\Big\{ 
\bItenGGG({M_C^2}, M_C^2, M_V^2) - 
\bItenGGG(M_C^2, {M_C^2}^T\!, M_V^2)
\Big\}~.  
} 
To arrive at this expression we exploited symmetries of the gauge
indices. Like diagram A, diagram H also contains derivatives $\der_m$
in the numerator, and therefore also this ``$\ominus$'' supergraphs
becomes a sum of ``$8$'' and ``$\ominus$'' scalar graphs. The
difference is that in this case they act on two different delta
functions. We make use of \eqref{scOminuspq} to obtain after some
algebra 
\equa{
i \gG^{\text{H}}_{2L} ~= & ~ \dsp 
\frac i8 \int \d^8 z\,  
f_{IK\, a} \, \bff_{JL\,}{}^{\ua}
\Big\{
2\, h\inv{}^{KL}\, \bJtenG(M^2, M_V^2) 
- G\inv{}^a{}_\ua\, \bJtenGG(M_V^2, M_V^2) + 
\non \\[2ex]& \dsp 
+ (T_M\gf)^a \, (\bgf T_N)_\ua\, 
\bItenGGG(M_V^2, M_V^2, M_C^2)
\Big\} 
+ \frac i8 \int \d^8 z\,  
\Big\{ 
 f_{IK\,b} (G\inv \bW)^{b \ua}\, 
\bff_{JL\,}{}^{\ub} ({G\inv}^T W)_{\ub a} + 
\non \\[2ex] & \dsp 
- f_{MK\, a} \, \bff_{NL\,}{}^{\ua}
\Big(
\gd^M{}_I (h\inv M_V^2)^N{}_J + 
\gd^N{}_J (h\inv M_V^2)^M{}_I
\Big) 
\Big\}
\bItenGG(M^2,M_V^2,M_V^2)
~. 
}
Diagram I becomes 
\equ{
i \gG^{\text{I}}_{2L} ~=~ \frac i2 \int \d^8 z\, 
f_{IK\, a}\, (M_C^2)_{JL}{}^{; \ua} \, 
\bItenGG (M^2,M_V^2,M_V^2)~. 
} 
Of course we also have the Hermitian conjugate of this graph. 
Finally, we have 
\equ{
i \gG^{\text{J}}_{2L} = 
i \int \d^8 z\, h_{LP}c^P{}_{IN}\, h_{JQ}c^Q{}_{KM}\, 
\bItenGGG({M_V^2}, M_V^2, M_V^2)~.  
} 
This complete our calculation of the contributions to the two loop \Kh\
potential, in the next and final subsection we collect the various
contributions together.

\subsection{Summary results for the effective \Kh\ potential at two loops}
\labl{sc:summary2L}

We collect and combine all contributions to the two
loop effective \Kh\ potential obtained in the previous subsections. 
These contributions result from the two loop graphs depicted in 
figures \ref{fg:8supergraph} and \ref{fg:0-supergraphs}. We use those
diagrams to refer the different contributions. The full two
loop corrections to the \Kh\ potential can be divided into two
parts depending on whether they are only present when the gauge
kinetic function is constant or not:  
\equ{
K_{2L} ~=~ 
K_{2L}^{\text{universal}} ~+~ 
K_{2L}^{\text{gauge kinetic}}~. 
\labl{full2LKh}
}
The part of the two loop correction that is present even when
the superpotential is constant we call ``universal'', and the part
that is only present for non--constant gauge kinetic functions we call
``gauge kinetic''. Before we below describe each of these contributions
separately, we would like to remind the reader that these results are
obtained for perpendicular $M_W^2$ and $M_G^2$, because we assumed
that the background equations of motion are satisfied. For field
configurations that do not, there can be additional corrections.

To write down the results in a compact way, we use the short hand
notations $\bJ$ and $\bI$ for the divergent integrals \eqref{sc8} and
\eqref{scOminus} containing matrix valued masses with the
subdivergences and poles removed. These integrals are evaluated in
appendix \ref{sc:2Lintegrals}, and their expressions are given in
\eqref{bJten} and \eqref{bIten}, respectively. As discussed in
appendix \ref{sc:massmatrices} it is in principle straightforward to
generalize the results for two loop scalar integral with scalar masses
to mass matrices. The only subtlety is that the function $\bIten$
takes a sightly different forms depending on the values of the masses
due to threshold effects: The functions $\gk(\bx_i)$ inside $\bI$, see
\eqref{bIten}, depend on whether the scalar quantity $(N^2)^2\,$,
defined in \eqref{flowcircle}, is positive or negative. As this is a
condition on scalar masses, to give an completely explicit expression
for $\bIten\,$,
one needs to go to a basis in which the mass matrices are diagonal. It
is therefore beneficial to be able to diagonalize the mass matrices 
explictly. (In the examples discussed in the next section we can
easily go to the diagonal basis to perform all computations.) 
Since we represent all the results here in terms of the integrals
$\bJten$ and $\bIten\,,$ this issue is not explicitly visible in our
expressions.

The universal part of the two loop \Kh\ potential takes the form 
\equ{
K_{2L}^{\text{universal}} ~=~ 
\frac 12 \, R^\ua{}_a{}^\ub{}_b ~ \bJten(M^2, M^2) 
~+~ 
\frac 16\, \bW^{;\ua\,\ub\,\uc} \,W_{;abc} ~ \bIten(M^2, M^2, M^2) ~+ 
\labl{2LKhuniversal}
\\[2ex] 
~+~ \frac 12 \, h_{LP}\,c^P{}_{IN}\, h_{JQ}\,c^Q{}_{KM}\, 
\Big\{ 
 \bItenGGG({M_C^2}, M_C^2, M_V^2) 
~-~ \bItenGGG(M_C^2, {M_C^2}^T, M_V^2)
\Big\}~+ 
\non \\[2ex] 
~+~ 
h_{LP}c^P{}_{IN}\, h_{JQ}c^Q{}_{KM}\, 
\bItenGGG({M_V^2}, M_V^2, M_V^2)
~-~ (G T_I \gf)^\ua{}_{;b} \, (\bgf T_J G)_a{}^{;\ub}\, 
\bItenG(M^2,M^2,M_V^2) 
~. 
\non 
}
The first term that is proportional to the curvature \eqref{curvature}
results from the ``$8$'' diagram combined with a part of diagram A. 
The next term depends on the triple covariant derivatives of the
superpotential \eqref{covderW}. It results from another part of
diagram A together with the diagrams B, C, $\overline{\text{C}}$ and
D.  The final part of diagram A cancels diagram F, which involve the
exchange of a vector multiplet, and therefore leaves no trace in
\eqref{2LKhuniversal}. The second line is a consequence of 
the ghost diagram G, and the third line is due to the graphs 
J and E in which a vector multiplet is exchanged.

This result is manifestly covariant under diffeomorphisms that
preserve the \Kh\ structure, because it is written in terms of the
curvature tensor \eqref{curvature} and the triple covariant derivative
of the superpotential \eqref{covderW}. The fact that the ``8''
supergraph in figure \ref{fg:8supergraph} and the supergraphs A to B
of figure \ref{fg:0-supergraphs} combine to covariant expressions
provides an important cross check of our two loop computation.  
Diagram D has been computed in refs.\ 
\cite{Buchbinder:1994xq,Buchbinder:1994iw,Buchbinder:1994df,Buchbinder:1996cy}
for the renormalizable Wess--Zumino model. 
The combination of the diagrams ``8'' and A-D have been computed in
ref.\ \cite{Buchbinder:2000ve,Petrov:1999qh} again for a single
ungauged chiral multiplet. (However, there seemed to be some
differences with our results, in particular that result is not covariant.) In
subsection \ref{sc:nonWZ} 
we discuss the non--renormalizable Wess--Zumino model, to make the
comparison with our results easier.

When the gauge kinetic function is not constant we find the additional
contributions 
\equ{
K_{2L}^{\text{gauge kinetic}}
~=~  \frac 18 \, 
f_{IK\, a} \, \bff_{JL\,}{}^{\ua}
\Big\{
2\, h\inv{}^{KL}\, \bJtenG(M^2, M_V^2) 
~-~ G\inv{}^a{}_\ua\, \bJtenGG(M_V^2, M_V^2) ~+~ 
\non \\[2ex] 
~+~ (T_M \gf)^a\, (\bgf T_N)_\ua \,\bItenGGG(M_V^2,M_V^2,M_C^2)
\Big\} 
~+~ \frac 18
\Big\{ 
f_{IK\,b} (G\inv \bW)^{b \ua}\, 
\bff_{JL\,}{}^{\ub} ({G\inv}^T W)_{\ub a} ~+ 
\non \\[2ex]
~-~ f_{MK\, a} \, \bff_{NL\,}{}^{\ua}
\Big(
\gd^M{}_I (h\inv M_V^2)^N{}_J ~+~ 
\gd^N{}_J (h\inv M_V^2)^M{}_I
\Big) 
\Big\} \, 
\bItenGG(M^2,M_V^2,M_V^2)
~+ 
\non \\[2ex]
~+~ \frac 12 \Big( 
f_{IK\, a}\, (M_C^2)_{JL\,}{}^{; \ua} ~+~ 
\bff_{IK\,}{}^{\ua}\, (M_C^2)_{JL\,; a} 
\Big) 
\bItenGG (M^2,M_V^2,M_V^2)~. 
\labl{2LKhgauge}
}
The terms that are proportional to the product of tensors $f$ and
$\bff$ arise from diagram H. The last line is the effect of diagram I
and it's Hermitian conjugate.


%% file: sc8graph.pstex_t
\begin{picture}(0,0)%
\includegraphics{sc8graph.pstex}%
\end{picture}%
\setlength{\unitlength}{4144sp}%
\begingroup\makeatletter\ifx\SetFigFont\undefined%
\gdef\SetFigFont#1#2#3#4#5{%
  \reset@font\fontsize{#1}{#2pt}%
  \fontfamily{#3}\fontseries{#4}\fontshape{#5}%
  \selectfont}%
\fi\endgroup%
\begin{picture}(1544,774)(2379,-5173)
\end{picture}

%% file: sc0-graph.pstex_t
\begin{picture}(0,0)%
\includegraphics{sc0-graph.pstex}%
\end{picture}%
\setlength{\unitlength}{4144sp}%
\begingroup\makeatletter\ifx\SetFigFont\undefined%
\gdef\SetFigFont#1#2#3#4#5{%
  \reset@font\fontsize{#1}{#2pt}%
  \fontfamily{#3}\fontseries{#4}\fontshape{#5}%
  \selectfont}%
\fi\endgroup%
\begin{picture}(1658,1652)(4794,-4487)
\end{picture}

%% file: scdoublegraph.pstex_t
\begin{picture}(0,0)%
\includegraphics{scdoublegraph.pstex}%
\end{picture}%
\setlength{\unitlength}{4144sp}%
\begingroup\makeatletter\ifx\SetFigFont\undefined%
\gdef\SetFigFont#1#2#3#4#5{%
  \reset@font\fontsize{#1}{#2pt}%
  \fontfamily{#3}\fontseries{#4}\fontshape{#5}%
  \selectfont}%
\fi\endgroup%
\begin{picture}(1893,732)(2200,-5152)
\end{picture}

%% file: 8supergraph.pstex_t
\begin{picture}(0,0)%
\includegraphics{8supergraph.pstex}%
\end{picture}%
\setlength{\unitlength}{4144sp}%
\begingroup\makeatletter\ifx\SetFigFont\undefined%
\gdef\SetFigFont#1#2#3#4#5{%
  \reset@font\fontsize{#1}{#2pt}%
  \fontfamily{#3}\fontseries{#4}\fontshape{#5}%
  \selectfont}%
\fi\endgroup%
\begin{picture}(1544,774)(2379,-5173)
\end{picture}

%% file: 0-superA.pstex_t
\begin{picture}(0,0)%
\includegraphics{0-superA.pstex}%
\end{picture}%
\setlength{\unitlength}{4144sp}%
\begingroup\makeatletter\ifx\SetFigFont\undefined%
\gdef\SetFigFont#1#2#3#4#5{%
  \reset@font\fontsize{#1}{#2pt}%
  \fontfamily{#3}\fontseries{#4}\fontshape{#5}%
  \selectfont}%
\fi\endgroup%
\begin{picture}(1781,1652)(4748,-4490)
\end{picture}

%% file: 0-superB.pstex_t
\begin{picture}(0,0)%
\includegraphics{0-superB.pstex}%
\end{picture}%
\setlength{\unitlength}{4144sp}%
\begingroup\makeatletter\ifx\SetFigFont\undefined%
\gdef\SetFigFont#1#2#3#4#5{%
  \reset@font\fontsize{#1}{#2pt}%
  \fontfamily{#3}\fontseries{#4}\fontshape{#5}%
  \selectfont}%
\fi\endgroup%
\begin{picture}(1786,1652)(4723,-4480)
\end{picture}

%% file: 0-superC.pstex_t
\begin{picture}(0,0)%
\includegraphics{0-superC.pstex}%
\end{picture}%
\setlength{\unitlength}{4144sp}%
\begingroup\makeatletter\ifx\SetFigFont\undefined%
\gdef\SetFigFont#1#2#3#4#5{%
  \reset@font\fontsize{#1}{#2pt}%
  \fontfamily{#3}\fontseries{#4}\fontshape{#5}%
  \selectfont}%
\fi\endgroup%
\begin{picture}(1782,1652)(4723,-4492)
\end{picture}

%% file: 0-superD.pstex_t
\begin{picture}(0,0)%
\includegraphics{0-superD.pstex}%
\end{picture}%
\setlength{\unitlength}{4144sp}%
\begingroup\makeatletter\ifx\SetFigFont\undefined%
\gdef\SetFigFont#1#2#3#4#5{%
  \reset@font\fontsize{#1}{#2pt}%
  \fontfamily{#3}\fontseries{#4}\fontshape{#5}%
  \selectfont}%
\fi\endgroup%
\begin{picture}(1761,1652)(4723,-4490)
\end{picture}

%% file: 0-superE.pstex_t
\begin{picture}(0,0)%
\includegraphics{0-superE.pstex}%
\end{picture}%
\setlength{\unitlength}{4144sp}%
\begingroup\makeatletter\ifx\SetFigFont\undefined%
\gdef\SetFigFont#1#2#3#4#5{%
  \reset@font\fontsize{#1}{#2pt}%
  \fontfamily{#3}\fontseries{#4}\fontshape{#5}%
  \selectfont}%
\fi\endgroup%
\begin{picture}(1806,1652)(4723,-4485)
\end{picture}

%% file: 0-superF.pstex_t
\begin{picture}(0,0)%
\includegraphics{0-superF.pstex}%
\end{picture}%
\setlength{\unitlength}{4144sp}%
\begingroup\makeatletter\ifx\SetFigFont\undefined%
\gdef\SetFigFont#1#2#3#4#5{%
  \reset@font\fontsize{#1}{#2pt}%
  \fontfamily{#3}\fontseries{#4}\fontshape{#5}%
  \selectfont}%
\fi\endgroup%
\begin{picture}(1781,1652)(4748,-4483)
\end{picture}

%% file: 0-superG.pstex_t
\begin{picture}(0,0)%
\includegraphics{0-superG.pstex}%
\end{picture}%
\setlength{\unitlength}{4144sp}%
\begingroup\makeatletter\ifx\SetFigFont\undefined%
\gdef\SetFigFont#1#2#3#4#5{%
  \reset@font\fontsize{#1}{#2pt}%
  \fontfamily{#3}\fontseries{#4}\fontshape{#5}%
  \selectfont}%
\fi\endgroup%
\begin{picture}(1672,1652)(4790,-4490)
\end{picture}

%% file: 0-superH.pstex_t
\begin{picture}(0,0)%
\includegraphics{0-superH.pstex}%
\end{picture}%
\setlength{\unitlength}{4144sp}%
\begingroup\makeatletter\ifx\SetFigFont\undefined%
\gdef\SetFigFont#1#2#3#4#5{%
  \reset@font\fontsize{#1}{#2pt}%
  \fontfamily{#3}\fontseries{#4}\fontshape{#5}%
  \selectfont}%
\fi\endgroup%
\begin{picture}(1929,1834)(4671,-4576)
\end{picture}

%% file: 0-superI.pstex_t
\begin{picture}(0,0)%
\includegraphics{0-superI.pstex}%
\end{picture}%
\setlength{\unitlength}{4144sp}%
\begingroup\makeatletter\ifx\SetFigFont\undefined%
\gdef\SetFigFont#1#2#3#4#5{%
  \reset@font\fontsize{#1}{#2pt}%
  \fontfamily{#3}\fontseries{#4}\fontshape{#5}%
  \selectfont}%
\fi\endgroup%
\begin{picture}(1929,1834)(4671,-4576)
\end{picture}

%% file: 0-superJ.pstex_t
\begin{picture}(0,0)%
\includegraphics{0-superJ.pstex}%
\end{picture}%
\setlength{\unitlength}{4144sp}%
\begingroup\makeatletter\ifx\SetFigFont\undefined%
\gdef\SetFigFont#1#2#3#4#5{%
  \reset@font\fontsize{#1}{#2pt}%
  \fontfamily{#3}\fontseries{#4}\fontshape{#5}%
  \selectfont}%
\fi\endgroup%
\begin{picture}(1929,1834)(4671,-4576)
\end{picture}%

%% file: examples.tex
\section{Simple applications}
\labl{sc:examples}

In this section we illustrate our general formulae for the effective
\Kh\ potential at one and two loops, which were given in \eqref{1LKh}
and \eqref{2LKhuniversal}, \eqref{2LKhgauge}, respectively, by
applying them to some simple supersymmetric models. 
As discussed in the summary section \ref{sc:summary2L}, the two loop
results are expressed in terms of the tensor integrals $\bJ$ and
$\bI\,$; their explicit forms can be found in appendix
\ref{sc:2Lintegrals}. In subsection \ref{sc:nonWZ} we consider as a 
first example a general non--renormalizable Wess--Zumino model, and
its simplification to the renormalizable version (most previous
investigations in the literature only consider this case). Our second
example, Super Quantum Electrodynamics, is discussed in subsection
\ref{sc:SQED}.

\subsection{The (non--)renormalizable Wess--Zumino model}
\labl{sc:nonWZ}

We consider a single chiral multiplet $\gf\,$, described by a \Kh\
potential $K ~=~ K(\bgf,\gf)$ and a superpotential $W(\gf)\,$. The
metric, connection and curvature 
read   
\equ{
G ~=~ K^{\ubar{1}}{}_1~, 
 \qquad 
\gG ~=~ G\inv K^{\ubar{1}}{}_{11}~, 
\qquad 
R ~=~ K^{\ubar{1}\,\ubar{1}}{}_{11} - \bgG\, G \, \gG~,
}
where $1$ and $\ubar{1}$ denote the differentiation w.r.t.\ $\gf$ and
$\bgf\,$, respectively. The triple covariant derivative of the
superpotential and the superpotential mass are given by 
\equ{
W_{;111} ~=~ W_{111} - 3 \, \gG\, W_{11}~, 
\qquad 
M_W^2 ~=~ G^{-2} \,| W_{11} |^2~. 
} 
Hence the one and two loop corrections to the effective \Kh\ potential
read 
\equ{
K_{1L} ~=~ 
\frac 1{16 \gp^2}\,\frac 12\,  M_W^2 
\Big( 2 ~-~ \ln \frac {M_W^2}{\bgm^2} \Big)~, 
\qquad 
K_{2L} ~= ~ 
\frac 12 \, R G^{-2}\, \bJ ~+~ 
\frac 16\, |W_{;111} |^2 G^{-3}\, \bI~, 
\labl{KHWZ}
}
with the short hand notations 
\equ{
\arry{rl}{
\bJ ~=\, & \dsp 
\frac1{(16 \gp^2)^2}\,  (M_W^2)^2
\Big( 1 - \ln \frac{M_W^2}{\bgm^2} \Big)^2~, 
\\[2ex] 
\bI ~=\, & \dsp 
\frac1{(16 \gp^2)^2}\, \frac 32\, M_W^2
\Big[ 
-5 + 4 \, \ln \frac{M_W^2}{\bgm^2} - \ln^2 \frac{M_W^2}{\bgm^2} 
+ 12 \, \gk(\bx)
\Big]~.  
}
}
As we discussed in summary subsection \ref{sc:summary2L} the form of
$\gk(\bx)$ depends on the sign of $(N^2)^2$ defined in
\ref{flowcircle}. Since $(N^2)^2 ~=~ -\frac 34 (M_W^2)^2$ is negative,
$\gk(\bx)$ is given by \eqref{gkN^2<0}. Using \eqref{shorts} we find
that $\bx ~=~ \frac 43 \sqrt 3\,$.

Of particular interest is the reduction to the renormalizable
Wess--Zumino model which we use in section \ref{sc:oneloop} to obtain
an independent cross check of our one loop result. In this case the
\Kh\ potential is trivial: $K ~=~ \bgf \gf\,$, which means that the
connection and curvature are all zero. The superpotential is given
by 
\equ{ 
W(\gf) ~=~ \frac 12\, m \, \gf^2 + \frac 1{3!}\, \gl\, \gf^3~,
} 
with $m$ and $\gl$ complex parameters, so that the mass 
$M_W^2 ~=~ |m + \gl \, \gf|^2\,.$ Hence the expressions for the
one and two loop \Kh\ potentials further simplify to 
\equ{
\arry{rl}{
K_{1L} ~=\, & \dsp 
\frac 1{16 \gp^2}\,\frac 12\,  M_W^2 
\Big( 2 ~-~ \ln \frac {M_W^2}{\bgm^2} \Big)~, 
\\[2ex] 
K_{2L} ~= \, & \dsp 
\frac1{(16 \gp^2)^2}\, \frac 14\, M_W^2
\Big\{ 
-5 + 4 \, \ln \frac{M_W^2}{\bgm^2} - \ln^2 \frac{M_W^2}{\bgm^2} 
+ 12 \, \gk(\bx)
\Big\}~.
}
\labl{KHrenWZ}
}

\subsection{Super Quantum Electrodynamics}
\labl{sc:SQED}

The theory of Super Quantum Electrodynamics consists of two oppositely
charged chiral multiplets $\gf_+$ and $\gf_-$ under a $\U{1}$ gauge
symmetry of which $V$ is the vector superfield. The \Kh\ potential for this model has the well known form 
\equ{
K ~= ~ 
\gf_+ e^{2V} \gf_+ + \gf_- e^{-2V} \gf_-~, 
\qquad 
W ~=~ 0~; 
}
we assume that super electron is massless to ensure that $M_W^2$ 
and  $M_G^2$ are perpendicular automatically. A Fayet--Iliopoulos term
can be included, but as this does not affect the results given below,
we have not done so here. The gauge kinetic action reads
\equ{
S_G ~=~ \frac 1{4g^2} \int \d^6 z\, \cW^\ga \cW_\ga~+~ \text{h.c.}~, 
} 
where $g^{-2} ~=~ h ~=~ f$ is the inverse gauge coupling.

Before we get to the explicit formulae for the one and two loop
results, we develop some properties of the masses matrices that appear
in those expressions. In the following it turns out to be convenient to
use the vector and matrix notation 
\equ{
\gf ~=~ \pmtrx{\gf_+ \\ \gf_-}~,
\qquad 
\bgf ~=~ \pmtrx{\bgf_+ & \bgf_-}~, 
\qquad 
T ~=~ \pmtrx{1 & 0 \\ 0 & -1}~, 
}
for the electron superfields and the charge operator $T\,$. Because the
theory is Abelian the vector and ghost mass parameters are equal
\equ{
M_V^2 ~=~ M_C^2 ~=~ 2 \, \bgf \gf~. 
} 
The superpotential mass is by definition zero and for Goldstone boson 
mass matrices we obtain 
\equ{
M_W^2 ~=~ 0~, 
\qquad 
M_G^2 ~=~ 2 g^2\, 
\pmtrx{\gf_+ \bgf_+ & - \gf_+ \bgf_- \\
-\gf_- \bgf_+ & \gf_- \bgf_-}
~=~ g^2 M_V^2 \, P_+~. 
}
The mass matrices can be expressed in term of the Hermitian projection
operators $P_+$ and $P_-$ which are mutual perpendicular. These
projectors  
\equ{
P_+ ~=~ \frac 1{\bgf \gf} 
\pmtrx{ \gf_+ \\ - \gf_-} \pmtrx{\bgf_+ & - \bgf_-}~, 
\qquad 
P_- ~=~ \frac 1{\bgf \gf} 
\pmtrx{ \bgf_- \\ \bgf_+} \pmtrx{\gf_- & \gf_+}~ 
} 
diagonalize the total mass matrix $M^2$ since 
\equ{
M^2 ~=~ M_W^2 + M_G^2 ~=~ m_+^2\, P_+ + m_-^2 \, P_-~, 
} 
with the mass eigenvalues $m_+^2 ~=~ g^2 M_V^2$ 
and $m_-^2 ~=~ 0\,.$ Because of the properties of these projection 
operators we can express the tensor $\bItenU(M^2,M^2,g^2M_V^2)\,$, for
the $\U{1}$ theory, as a sum of scalar integrals $\bI\,$:  
\equ{
\bItenU(M^2, M^2, g^2 M_V^2) ~=~ 
\sum_{r,s} \, (P_r)^a{}_\ua \, (P_s)^b{}_\ub\, 
\bI(m_r^2, m_s^2, g^2 M_V^2)~, 
}
where the sum is over $r,s = +, -\,.$ This shows that the evaluation of
the two loop result reduces computing the traces $\tr( P_r T P_s T)$
because $(GT\gf)^\ua{}_{;a} ~=~ T^\ua{}_a\,,$ and  
$(\bgf T G)_b{}^{;\ub} ~=~ T^\ub{}_b\,.$ Explicitly we find for these
traces 
\equ{
\tr(P_+ T P_+ T) ~=~ \tr(P_- T P_- T) ~=~ 
\Big( \frac{\bgf \gs_3 \gf}{\bgf \gf}\Big)^2~, 
\qquad 
\tr(P_+ T P_- T) ~=~ 
\Big|  \frac {\gf^T \gs_1 \gf}{\bgf \gf} \Big|^2~, 
} 
with $\gs_1$ and $\gs_3$ the standard Pauli matrices.

After this exposition it is not difficult to see that the one and two
loop corrections to the effective \Kh\ potential are given by the
following expressions. At the one loop level we find 
\equ{
K_{1L} ~=~ - \frac 1{16\gp^2}\,{g^2 M_V^2}
\Big( 
2 - \ln \frac {g^2 M_V^2}{\bgm^2}
\Big)~. 
}
We have dropped the contribution coming from the superpotential
mass matrix as it gives a mere constant which is irrelevant for the
\Kh\ geometry. The two loop result takes the form 
\equ{
K_{2L} = 
- \Big\{ \bI(m_+^2, m_+^2, g^2 M_V^2) + 
\bI(m_-^2, m_-^2, g^2 M_V^2) \Big\}
\Big( \frac{\bgf \gs_3 \gf}{\bgf \gf} \Big)^2 
-2\, \bI(m_+^2, m_-^2, g^2 M_V^2) 
\Big| \frac {\gf^T \gs_1 \gf}{\bgf \gf} \Big|^2~. 
}


%% file: 1Lintegrals.tex
\section{One loop scalar integrals}
\labl{sc:1Lintegrals}

This appendix is devoted to the evaluation of some one loop scalar
integrals that arise in the main text of this paper. We compute these
scalar integrals in the \MSbar\ scheme: We evaluate the integrals in  
$D ~=~ 4 - 2 \ge$ dimensions, and we introduce the renormalization
scale $\gm$ such that all $D$ dimensional integrals have the same mass
dimensions as their divergent four dimensional counter parts. Even
though these one loop integrals are well--known, we feel that 
 they should be collected here, because their importance for the
 subtraction of the subdivergences at the two loop level. 
Moreover, this allows us to introduce some notation for renormalized
quantities that we are employing throughout this work.

At the one loop level we encounter three different types of
integrals. The first integral reads
\equ{
J(m^2) ~=~ \int \frac{\d^D p}{(2\pi)^D \gm^{D-4}} \, 
\frac{1}{p^2 + m^2} 
~=~ - \frac {m^2}{16 \pi^2} \, 
\Big( 4\gp \frac {\gm^2}{m^2}  \Big)^{2- \frac D2}
\, \frac{\gG(2 - \frac D2)}{\frac D2 -1}
~. 
\labl{J1Lint}
}
For the applications to two loop graphs we need the expansion of this
integral to first order in $\ge\,$: 
\equ{
J(m^2) ~=~ - \frac {m^2}{16 \pi^2} \, 
\Big[  
\frac 1\ge + 1 - \ln \frac{m^2}{\bgm^2} + 
\ge \Big( 
1 + \frac 12 \gz(2) - \ln \frac{m^2}{\bgm^2} 
+ \frac 12 \ln^2 \frac{m^2}{\bgm^2} 
\Big) 
\Big]~. 
\labl{J1Lexp}
} 
Here we have introduced the \MSbar\ scale 
\(
\bgm^2 ~=~ 4 \gp e^{-\gg} \gm^2
\) 
with the Euler constant $\gg$ and $\gz(2) ~=~ \frac {\gp^2}6\,.$
The other two integrals 
\equ{
L(m^2) ~=~ \int \frac{\d^D p}{(2\pi)^D \gm^{D-4}} \, 
\frac{1}{p^2} \ln\Big({1 + \frac{m^2}{p^2}}\Big)  
~=~  \frac {m^2}{16 \pi^2} \, 
\Big( 4\gp \frac {\gm^2}{m^2}  \Big)^{2- \frac D2}
\, \frac{\gG(2 - \frac D2)}{( \frac D2 -1)^2}
\labl{Log1Lint}
}
and 
\equ{
S(m^2) ~=~ \int \frac{\d^D p}{(2\pi)^D \gm^{D-4}} \, 
\frac{1}{(p^2 + m^2)^2} 
~=~  \frac {1}{16 \pi^2} \, 
\Big( 4\gp \frac {\gm^2}{m^2}  \Big)^{2- \frac D2}
\gG(2 - \sfrac D2) 
}
are only relevant for one loop computations in this work, therefore we
only need to expand them to zeroth order in $\ge\,$: 
\equ{
L(m^2) ~=~ \frac {m^2}{16 \pi^2} \, 
\Big[  
\frac 1\ge + 2 - \ln \frac{m^2}{\bgm^2} 
\Big]~, 
\qquad 
S(m^2) ~=~ \frac 1{16\pi^2}\, \Big[
\frac 1\ge  - \ln \frac{m^2}{\bgm^2}
\Big]~. 
\labl{Log1LintRes} 
}


%% file: 2Lintegrals.tex
\section{Two loop scalar integrals}
\labl{sc:2Lintegrals}

At the two loop level we encounter two different scalar momentum
integrals. As can be seen in figure \ref{fg:2Lgraphs} they have the
topology of  an ``$8$'' (figure \ref{fg:2Lgraphs}.a) and ``$\ominus$''  
(figure \ref{fg:2Lgraphs}.b). (The third topology depicted in figure 
\ref{fg:2Lgraphs}.c is the same as figure \ref{fg:2Lgraphs}.a as a
momentum integral, and can therefore be disregarded in this appendix.)
Being two loop graphs, these integrals contain subdivergences. The
subtraction of subdivergences is crucial, because otherwise one would
not obtain local counter terms from the two loop level onwards. A
generic feature of such non--local counter terms is, that they involve
terms like $\frac 1\ge \, \ln \bgm^2\,$. In our calculation such terms 
are also absent after all subdivergences are subtracted off. In
this work we take the practical approach that these subdivergences can
be subtracted off on a diagram by diagram level directly, instead of 
computing explicitly one loop graphs with one loop counter terms
inserted. The expression of a  two loop integral $I$ with all its
subdivergences subtracted off is denoted by $\hat I\,$.

\subsection{The scalar ``$8$'' integral}
\labl{sc:scalar8graph}

The ``8'' graph is easy to evaluate as it is the product of two one
loop integrals defined in \eqref{J1Lint}: 
\equ{
J(m_1^2, m_2^2) ~=~ J(m_1^2)\, J(m_2^2)~. 
\labl{sc8}
} 
The subtraction of the subdivergences leads to 
\equ{
\hat J(m_1^2,m_2^2) ~=~ J(m_1^2, m_2^2) + 
\frac 1{16\gp^2}\, \frac 1\ge\, \Big( 
m_2^2\, J(m_1^2)+ m_1^2 \, J(m_2^2)
\Big)~. 
} 
Expanding this to zeroth order in $\ge$ gives 
\equ{
\hat J(m_1^2,m_2^2) ~=~  \frac {m_1^2 m_2^2}{(16\gp^2)^2}\, 
\Big[ 
- \frac 1{\ge^2} + 
\Big( 1 -\ln  \frac {m_1^2}{\bgm^2} \Big) 
\Big( 1 - \ln \frac {m_2^2}{\bgm^2} \Big) 
\Big]~. 
\labl{Jhat}
}
We refer to this expression as $\bJ(m_1^2, m_2^2)\,$, when the pole 
in $\ge^2$ is subtracted off.

\subsection{The basic scalar ``$\boldsymbol{\ominus}$'' diagram}
\labl{sc:scalar0-graph}

The evaluation of the ``$\ominus$'' graph is much more involved when
all three propagator lines correspond to three different masses
$m_1^2, m_2^2$ and $m_3^2\,$: 
\equ{
I(m_1^2, m_2^2, m_3^2) ~=~ 
\int \frac{\d^D p\,  \d^D q}{(2\pi)^{2D} \gm^{2(D-4)}} \, 
\frac{1}{p^2 + m_1^2} \, 
\frac{1}{q^2 + m_2^2} \, 
\frac{1}{(p+q)^2 + m_3^2} 
~. 
\labl{scOminus} 
} 
To compute this integral directly is difficult; various methods to do
so can be found in the literature \cite{Fogleman:1983hm,Fogleman:1984eh,Miller:1983ci,Davydychev:1992mt,Davydychev:1993pg,Misiak:1994zw}.  
These results generically lead to complicated expressions, or are
only valid in specific limits or certain simplifying assumptions.
In particular, the fact that  this integral is manifestly symmetric in
the masses is generically lost. An elegant indirect way of computing
this integral leading to a surprisingly simple result has been
presented in refs.\ \cite{Ford:1991hw,Ford:1992pn}. (See also 
\cite{Martin:2001vx}.) For completeness
we review their method of characteristics.

The essential idea of the method of characteristics is, instead of
computing the integral \eqref{scOminus} head on, to relate the
expression of this integral for various values of the masses to each
other. In this way the complicated initial integral can be expressed
in terms of a combination of simpler integrals. In particular, when
two of the three masses are zero, the evaluation of the integral can
be performed by standard methods directly:  
\equ{
I(m^2, 0, 0) ~=~ 
\frac {m^2}{(16 \gp^2)^2} \, 
\Big( 4\gp \frac {\gm^2}{m^2}  \Big)^{4- D} \, 
\frac{\gG(3-D) \gG(\frac D2 -1)^2 \gG(2-\frac D2)}
{\gG(\frac D2)}~. 
\labl{easyint}
} 
Therefore we seek to relate $I(m_1^2,m_2^2,m_3^2)$ to 
$I(M_1^2, M_2^2,0)$ and then that one to $I(2 N^2, 0, 0)\,$. (The
masses $M_1^2, M_2^2$ and $N^2$ will be determined below.)

To related the expressions of integral \eqref{scOminus} for different
values of $m_1^2, m_2^2$ and $m_3^2$ to each other means that we
would like to describe a flow through the space of these masses. Such
a flow can be encoded by a partial differential that the integral
\eqref{scOminus} satisfies. Of course this equation should be
sufficiently simple to be of practical use. For the case at hand a
convenient partial differential equation is given by 
\equ{
\Big[ 
(m_1^2 - m_2^2) \pp{m_3^2} + 
\text{cycl.}
\Big] I(m_1^2,m_2^2,m_3^2)  ~=~ 
\pp[J(m_3^2)]{m_3^2} 
\Big( J(m_1^2) - J(m_2^2) \Big) + \text{cycl.}~,
\labl{PDE}
} 
where ``$+\text{cycl.}$'' indicates the summation over the cyclic
permutation of the labels $1,2,3$ on the masses. This partial
differential equation is obtained by combining various equations that
can be found by integrating the total derivative 
\equ{
\pp{p^\gm} \Big[
{p^\gm} \, 
\frac{1}{p^2 + m_1^2} \, 
\frac{1}{q^2 + m_2^2} \, 
\frac{1}{(p+q)^2 + m_3^2} 
\Big] 
} 
over $p$ and $q$, and similar for $\pp{q^\gm}[q^\gm
\ldots]\,$. Rewriting the result as partial differentiation w.r.t.\
the masses, and using cyclic permutations of the labels on the
masses. Along the lines of flow, described by the equations   
\equ{
\dd[m_1^2]{t} ~=~ m_2^2 - m_3^2~, 
\qquad 
\dd[m_2^2]{t} ~=~ m_3^2 - m_1^2~, 
\qquad 
\dd[m_3^2]{t} ~=~ m_1^2 - m_2^2~, 
\labl{flow} 
} 
this partial differential equation can be integrated. Such a flow line is
an arc of a circle defined by the intersection of a plane and a sphere 
\equ{
m_1^2 + m_2^2 + m_3^2 ~=~ \bm^2~, 
\qquad 
(m_1^2)^2 + (m_2^2)^2 + (m_3^2)^2 ~=~  
2 (N^2)^2 + \frac 12 (\bm^2)^2~. 
\labl{flowcircle}
}
It follows from the flow equations \eqref{flow} that $\bm^2$
and $N^2$ are constants. If $(N^2)^2 \geq 0\,$, it follows that the 
point $(M_1^2, M_2^2, M_3^2=0)\,$, with
\equ{
M_1^2 ~=~ \frac{\bm^2}2 + N^2~, 
\qquad 
M_2^2 ~=~ \frac{\bm^2}2 - N^2~, 
}
lies on the same circle as the starting point $(m_1^2,m_2^2,m_3^2)\,$, 
and therefore we can obtain a relation between 
$I(m_1^2,m_2^2,m_3^2)$ and $I(M_1^2, M_2^2,0)$ by integrating the
partial differential equation \eqref{PDE} along this arc. Similarly,
we can consider the partial differential equation 
\equ{
(M_2^2 - M_1^2) \Big( \pp{M_1^2} + \pp{M_2^2} \Big) 
I(M_1^2, M_2^2, 0) ~=~  
\pp[J(M_2^2)]{M_2^2} J(M_1^2) - 
\pp[J(M_1^2)]{M_1^2} J(M_2^2)~ 
} 
for $I(M_1^2, M_2^2,0)\,$. In this case the flow is along straight lines,
because the right hand side of this equation is also proportional to 
$(M_2^2 - M_1^2)\,.$ Therefore we can relate the integrals
$I(M_1^2,M_2^2,0)$ and  $I(2N^2, 0, 0)\,$, with 
$2N^2 = M_1^2 - M_2^2$. By combining these results, the expression
\eqref{easyint}, and using the short--hand notations 
\equ{
\bm_i^2 ~= \frac{\bm^2}2 - m_i^2~, 
\qquad 
\bx_i ~=~ \frac {\bm_i^2}{|N|^2}~, 
\labl{shorts}
} 
one can obtain the following factorized form 
\equ{ 
I(m_1^2,m_2^2,m_3^2) ~=~ - I_C \, 
\Big[ 
F(\bx_2) + F(\bx_3) - F(-\bx_1) 
- \frac 12 \frac{\gG(\frac 32 - \frac D2) \gG(\frac D2 -1)}{\gG(\frac 12)}
\Big]~, 
\labl{Factorized}
}
where we have identified a common factor 
\equ{
I_C ~=~  \frac {N^2}{(16 \gp^2)^2} \, 
\Big( 4\gp \frac {\gm^2}{N^2} \Big)^{4-D}\, 
\frac {\gG(2 - \frac D2)^2}{\frac D2 - 1}~, 
\quad\text{and}\quad 
F(\bx) ~=~ \int_1^\bx \d s\, (s^2 - 1)^{\frac D2-2}~
}
is a shorthand for the remaining integral. The expression
\eqref{Factorized} is not manifestly symmetric in all permutations of 
$\bx_1, \bx_2, \bx_3\,$. But this symmetry is present because 
$F(x) - F(-y) ~=~ F(y)- F(-x)$ for any $x,y\,$. In the ``$\ominus$''
diagram one can identify three one loop subgraphs, it follows that the
subdivergences are removed by defining  
\equ{
\hat I (m_1^2,m_2^2,m_3^2) ~=~ 
I(m_1^2,m_2^2,m_3^2) - \frac 1{16\gp^2} \, \frac 1 \ge \, 
\Big( J(m_1^2) + J(m_2^2) + J(m_3^2) \Big)~, 
\labl{Ominus}
} 
with $J(m^2)$ given in \eqref{J1Lexp}.

To expand this expression to the zeroth order in $\ge$ explicitly, we
need to compute various integrals. Most of them are straightforward,
except 
\equ{
\int_1^\bx \d s\, \ln|s + 1| \,\ln | s - 1| ~=~ 
\frac 12 \ln^22 + 2 \ln 2 - 2 + 2 \bx - \bx \, \ln | \bx^2 -1| +  
\ln \Big| \frac {\bx-1}{\bx+1} \Big| + 
\qquad \qquad 
\\[2ex] 
\qquad \qquad + \bx \, \ln|\bx + 1|\,  \ln|\bx -1| 
+ \frac 12\, \ln|\bx^2-1|\,  \ln \Big| \frac {\bx-1}{\bx+1} \Big| 
- 4 \gk(\bx)~, 
\non } 
with 
\equ{
\gk(\bx) ~=~  - \int_0^{a} \d t\, 
\ln | \sinh t |
~- \frac 3{16} \, \ln^2 2 - \frac 14\, \gz(2)~, 
\qquad 
a ~=~ \coth\inv \bx~. 
\labl{gkN^2>0}
}
After some algebra we can express the ``$\ominus$'' graph with the 
subdivergences subtracted as   
\equa{
\hat I (m_1^2, m_2^2, m_3^2) ~= & \, \dsp
 \frac 12 \, \frac 1{(16 \gp^2)^2} 
\Big[
\bm^2 \Big( \frac 1{\ge^2} - \frac 1\ge \Big) - 5 \, \bm^2 
+ 4 \, \sum_i m_i^2 \,  \ln \frac {m_i^2}{\bgm^2} + 
\labl{hatIint}
\\[2ex] &  \dsp 
-  2 \Big( 
\bm_1^2 \, \ln \frac {m_2^2}{\bgm^2} \, \ln \frac {m_3^2}{\bgm^2}
+ \text{cycl.} 
\Big) 
+ 8\, N^2 \Big( 
- \gk(-\bx_1) + \gk(\bx_2) + \gk(\bx_3)
\Big) 
\Big]~, 
\non 
} 
where $\bx_i$ is given in \eqref{shorts}. 
The calculation that we have reviewed here is valid for $(N^2)^2 \geq 0\,$,
in the opposite case one can show \cite{Ford:1992pn} that
\eqref{hatIint} still holds, but now with $N^2 ~\ra~ |N|^2$
everywhere even when it appears implicit, $- \gk(-\bx_1) ~\ra~ \gk(\bx_1)$
and   
\equ{
\gk(\bx)  ~=~ -\int_0^a \d t \, \ln|\cos t|~, 
\qquad 
a ~=~ \tan\inv \bx~. 
\labl{gkN^2<0}
}

\subsection{Other scalar ``$\ominus$'' diagrams}
\labl{sc:other0-}

In the computation of the supergraphs in section \ref{sc:ominusgraphs}
we encounter integrals that are the same as \eqref{scOminus} except that
the numerator is more complicated. Fortunately, these integrals can be
reduced to \eqref{scOminus} and \eqref{sc8}. In particular, we have
that: 
\equ{
\int \frac{\d^D p\,  \d^D q}{(2\pi)^{2D} \gm^{2(D-4)}} \, 
\frac{p^2}{p^2 + m_1^2} \, 
\frac{1}{q^2 + m_2^2} \, 
\frac{1}{(p+q)^2 + m_3^2} 
~=~ J(m_2^2, m_3^2) - m_1^2 \, I(m_1^2, m_2^2, m_3^2)~, 
\labl{scOminusp^2}
}
and 
\equ{
\int \frac{\d^D p\,  \d^D q}{(2\pi)^{2D} \gm^{2(D-4)}} \, 
\frac{2 \, p \cdot q}{(p^2 + m_1^2)(q^2 + m_2^2)} \, 
\frac{1}{(p+q)^2 + m_3^2} 
~=~ - J(m_2^2, m_3^2) - J(m_1^2,m_3^2) + J(m_1^2, m_2^2)
\non \\[2ex]  
+ (m_1^2 + m_2^2 - m_3^2)\, I(m_1^2, m_2^2, m_3^2)~. 
\labl{scOminuspq}
}

\subsection{Generalization to mass matrices}
\labl{sc:massmatrices}

In the section \ref{sc:twoloop} we are dealing with the two loop
calculation of supersymmetric theories with multiple scalar and vector
multiplets, and therefore the masses are matrices rather than simple
numbers. Here we define our notation to describe two loop
integrals that involve mass matrices with all subdivergences and the
$1/\ge$ poles subtracted. For notational convenience we choose to
include the inverse metric $G\inv$ in these definitions, so that the
expressions contain an equal number of holomorphic and
anti--holomorphic indices.

For the integral $J\,$, given in \eqref{sc8}, we obtain a factorized form 
\equ{
\bJten(m_1^2, m_2^2) ~=~ 
 \frac {1}{(16\gp^2)^2}\, 
\Big[ 
m_1^2 
\Big( 1 -\ln  \frac {m_1^2}{\bgm^2} \Big)  G\inv 
\Big]^a_{~\, \ua}
\, 
\Big[ m_2^2 
\Big( 1 - \ln \frac {m_2^2}{\bgm^2} \Big) G\inv 
\Big]^b_{~\, \ub} 
~. 
\labl{bJten}
}
The expression for the integral $I\,$, defined in \eqref{scOminus}, with
the subdivergences and poles subtracted does not factorize, see
\eqref{hatIint}, consequently the matrix generalization is more
complicated:  
\equ{
\arry{l}{ \dsp 
\bIten(m_1^2,m_2^2,m_3^2) ~= \,   
\frac 12 \, \frac1{(16\gp^2)^2} \, 
\Big\{
\Big( 
- \frac 52 \, m_1^2 + 4 \, m_1^2 \, \ln \frac{m_1^2}{\bgm^2} G\inv
\Big)^a_{~\, \ua}  G\inv{}^b_{~\, \ub}\, G\inv{}^c_{~\, \uc}  + 
\\[2ex] \dsp \qquad 
- {G\inv}^a_{~\,\ua} 
\Big( m_2^2\, \ln \frac{m_2^2}{\bgm^2} \, G\inv \Big)^b_{~\, \ub} 
\Big( \ln \frac{m_3^2}{\bgm^2} \, G\inv \Big)^c_{~\, \uc} 
- {G\inv}^a_{~\,\ua} 
\Big(\ln \frac{m_2^2}{\bgm^2} \, G\inv \Big)^b_{~\, \ub} 
\Big(m_3^2 \, \ln \frac{m_3^2}{\bgm^2} \, G\inv \Big)^c_{~\, \uc} + 
}
\non \\
+ (m_1^2\, G\inv)^a_{~\, \ua} 
\Big( \ln \frac{m_2^2}{\bgm^2}\, G\inv \Big)^b_{~\, \ub} 
\Big( \ln \frac{m_3^2}{\bgm^2} \, G\inv \Big)^c_{~\, \uc} 
+ \text{cycl.} 
+ \text{etc.} 
\Big\}~. 
\labl{bIten}
} 
Here the ``$+\text{cycl.}$'' denote the cyclic permutation of the
labels $1,2,3$ and the corresponding indices $a,b,c$ and
$\ua,\ub,\uc\,$, and with ``$+\text{etc.}$'' we refer to the expansion
of the $N^2\, \sum_i \gk(\bx_i)$ term in \eqref{hatIint}. 
The functions $\gk(\bx_i)$ take two different forms 
(\eqref{gkN^2>0} or \eqref{gkN^2<0}) depending on in which regime 
($(N^2) > 0$ or $(N^2)^2 < 0\,$) they are evaluated. Since this
condition is written down for scalar masses, it means that in the case
matrices it has to be evaluated in the diagonal basis.

This notation can be extended to included also adjoint indices $I, J,
\ldots$. For those we take the convention that all indices are written
as superscripts, i.e.\ these expressions include appropriate powers of
the inverse metric $h\inv$ defined below \eqref{GFfunction}. 
The integrals given in appendix \ref{sc:other0-} can be 
reinterpreted in an analogous way as matrix expressions.


%% file: paper.bbl
\providecommand{\href}[2]{#2}\begingroup\raggedright\begin{thebibliography}{10}

\bibitem{Jackiw:1974cv}
R.~Jackiw ``Functional evaluation of the effective potential'' {\em Phys. Rev.}
  {\bf D9} (1974)
1686.

\bibitem{vonGersdorff:2005ce}
G.~von Gersdorff and A.~Hebecker ``Radius stabilization by two-loop casimir
  energy'' {\em Nucl. Phys.} {\bf B720} (2005) 211--227
\href{http://www.arXiv.org/abs/hep-th/0504002}{[{\tt hep-th/0504002}]}.

\bibitem{Grisaru:1979wc}
M.~T. Grisaru, W.~Siegel, and M.~Rocek ``Improved methods for supergraphs''
  {\em Nucl. Phys.} {\bf B159} (1979)
429.

\bibitem{Novikov:1982px}
V.~A. Novikov, M.~A. Shifman, A.~I. Vainshtein, and V.~I. Zakharov ``Instantons
  in supersymmetric theories'' {\em Nucl. Phys.} {\bf B223} (1983)
445.

\bibitem{Shifman:1986zi}
M.~A. Shifman and A.~I. Vainshtein ``Solution of the anomaly puzzle in {SUSY}
  gauge theories and the {W}ilson operator expansion'' {\em Nucl. Phys.} {\bf
  B277} (1986)
456.

\bibitem{Seiberg:1993vc}
N.~Seiberg ``Naturalness versus supersymmetric nonrenormalization theorems''
  {\em Phys. Lett.} {\bf B318} (1993) 469--475
\href{http://www.arXiv.org/abs/hep-ph/9309335}{[{\tt hep-ph/9309335}]}.

\bibitem{Seiberg:1994bz}
N.~Seiberg ``Exact results on the space of vacua of four-dimensional {SUSY}
  gauge theories'' {\em Phys. Rev.} {\bf D49} (1994) 6857--6863
\href{http://www.arXiv.org/abs/hep-th/9402044}{[{\tt hep-th/9402044}]}.

\bibitem{Intriligator:1994jr}
K.~A. Intriligator, R.~G. Leigh, and N.~Seiberg ``Exact superpotentials in
  four-dimensions'' {\em Phys. Rev.} {\bf D50} (1994) 1092--1104
\href{http://www.arXiv.org/abs/hep-th/9403198}{[{\tt hep-th/9403198}]}.

\bibitem{Seiberg:1994rs}
N.~Seiberg and E.~Witten ``Electric - magnetic duality, monopole condensation,
  and confinement in {N=2} supersymmetric {Y}ang-{M}ills theory'' {\em Nucl.
  Phys.} {\bf B426} (1994) 19--52
\href{http://www.arXiv.org/abs/hep-th/9407087}{[{\tt hep-th/9407087}]}.

\bibitem{Grisaru:1996ve}
M.~T. Grisaru, M.~Rocek, and R.~von Unge ``Effective {K\"ahler} potentials''
  {\em Phys. Lett.} {\bf B383} (1996) 415--421
\href{http://www.arXiv.org/abs/hep-th/9605149}{[{\tt hep-th/9605149}]}.

\bibitem{deWit:1996kc}
B.~de~Wit, M.~T. Grisaru, and M.~Rocek ``Nonholomorphic corrections to the
  one-loop {N=2} super {Yang- Mills} action'' {\em Phys. Lett.} {\bf B374}
  (1996) 297--303
\href{http://www.arXiv.org/abs/hep-th/9601115}{[{\tt hep-th/9601115}]}.

\bibitem{Pickering:1996gt}
A.~Pickering and P.~C. West ``The one loop effective super-potential and non-
  holomorphicity'' {\em Phys. Lett.} {\bf B383} (1996) 54--62
\href{http://www.arXiv.org/abs/hep-th/9604147}{[{\tt hep-th/9604147}]}.

\bibitem{Buchbinder:1999jw}
I.~L. Buchbinder, M.~Cvetic, and A.~Y. Petrov ``One-loop effective potential of
  {N = 1} supersymmetric theory and decoupling effects'' {\em Nucl. Phys.} {\bf
  B571} (2000) 358--418
\href{http://www.arXiv.org/abs/hep-th/9906141}{[{\tt hep-th/9906141}]}.

\bibitem{Brignole:2000kg}
A.~Brignole ``One-loop kaehler potential in non-renormalizable theories'' {\em
  Nucl. Phys.} {\bf B579} (2000) 101--116
\href{http://www.arXiv.org/abs/hep-th/0001121}{[{\tt hep-th/0001121}]}.

\bibitem{Gaillard:1996hs}
M.~K. Gaillard, V.~Jain, and K.~Saririan ``Supergravity coupled to chiral and
  {Y}ang-{M}ills matter at one loop'' {\em Phys. Lett.} {\bf B387} (1996)
  520--528
\href{http://www.arXiv.org/abs/hep-th/9606135}{[{\tt hep-th/9606135}]}.

\bibitem{Gaillard:1994mn}
M.~K. Gaillard ``Pauli-{V}illars regularization of globally supersymmetric
  theories'' {\em Phys. Lett.} {\bf B347} (1995) 284--290
\href{http://www.arXiv.org/abs/hep-th/9412125}{[{\tt hep-th/9412125}]}.

\bibitem{Gaillard:1994sf}
M.~K. Gaillard ``Pauli-{V}illars regularization of supergravity coupled to
  chiral and {Y}ang-{M}ills matter'' {\em Phys. Lett.} {\bf B342} (1995)
  125--131
\href{http://www.arXiv.org/abs/hep-th/9408149}{[{\tt hep-th/9408149}]}.

\bibitem{Falkowski:2005fm}
A.~Falkowski ``On the one-loop kaehler potential in five-dimensional
  brane-world supergravity'' {\em JHEP} {\bf 05} (2005) 073
\href{http://www.arXiv.org/abs/hep-th/0502072}{[{\tt hep-th/0502072}]}.

\bibitem{Buchbinder:1994xq}
I.~L. Buchbinder, S.~M. Kuzenko, and A.~Y. Petrov ``Superfield chiral effective
  potential'' {\em Phys. Lett.} {\bf B321} (1994)
372--377.

\bibitem{Buchbinder:1994iw}
I.~L. Buchbinder, S.~Kuzenko, and Z.~Yarevskaya ``Supersymmetric effective
  potential: Superfield approach'' {\em Nucl. Phys.} {\bf B411} (1994)
665--692.

\bibitem{Buchbinder:1994df}
I.~L. Buchbinder, S.~M. Kuzenko, A.~Y. Petrov, and Z.~V. Yarevskaya
  ``Superfield effective potential''
\href{http://www.arXiv.org/abs/hep-th/9501047}{[{\tt hep-th/9501047}]}.

\bibitem{Buchbinder:1996cy}
I.~L. Buchbinder, S.~M. Kuzenko, and A.~Y. Petrov ``Superfield effective
  potential in the two loop approximation'' {\em Phys. Atom. Nucl.} {\bf 59}
  (1996)
148--153.

\bibitem{Buchbinder:2000ve}
I.~L. Buchbinder and A.~Y. Petrov ``Superfield effective action within the
  general chiral superfield model'' {\em Phys. Atom. Nucl.} {\bf 63} (2000)
1657--1670.

\bibitem{Petrov:1999qh}
A.~Y. Petrov ``Effective action in general chiral superfield model''
\href{http://www.arXiv.org/abs/hep-th/0002013}{[{\tt hep-th/0002013}]}.

\bibitem{Buchbinder:1998qv}
I.~L. Buchbinder and S.~M. Kuzenko ``Ideas and methods of supersymmetry and
  supergravity: {Or} a walk through superspace''. Bristol, UK: IOP (1998) 656
  p.

\bibitem{Wess:1992cp}
J.~Wess and J.~Bagger ``Supersymmetry and supergravity''. Princeton, USA: Univ.
  Pr. (1992) 259 p.

\bibitem{Gates:1983nr}
S.~J. Gates, M.~T. Grisaru, M.~Rocek, and W.~Siegel ``Superspace, or one
  thousand and one lessons in supersymmetry'' {\em Front. Phys.} {\bf 58}
  (1983) 1--548
\href{http://www.arXiv.org/abs/hep-th/0108200}{[{\tt hep-th/0108200}]}.

\bibitem{West:1990tg}
P.~C. West ``Introduction to supersymmetry and supergravity''. Singapore,
  Singapore: World Scientific (1990) 425 p.

\bibitem{Siegel:1979wq}
W.~Siegel ``Supersymmetric dimensional regularization via dimensional
  reduction'' {\em Phys. Lett.} {\bf B84} (1979)
193.

\bibitem{Capper:1979ns}
D.~M. Capper, D.~R.~T. Jones, and P.~van Nieuwenhuizen ``Regularization by
  dimensional reduction of supersymmetric and nonsupersymmetric gauge
  theories'' {\em Nucl. Phys.} {\bf B167} (1980)
479.

\bibitem{Jack:1994bn}
I.~Jack, D.~R.~T. Jones, and K.~L. Roberts ``Equivalence of dimensional
  reduction and dimensional regularization'' {\em Z. Phys.} {\bf C63} (1994)
  151--160
\href{http://www.arXiv.org/abs/hep-ph/9401349}{[{\tt hep-ph/9401349}]}.

\bibitem{Jack:1997sr}
I.~Jack and D.~R.~T. Jones ``Regularisation of supersymmetric theories''
\href{http://www.arXiv.org/abs/hep-ph/9707278}{[{\tt hep-ph/9707278}]}.

\bibitem{Bagger:1995ay}
J.~Bagger, E.~Poppitz, and L.~Randall ``Destabilizing divergences in
  supergravity theories at two loops'' {\em Nucl. Phys.} {\bf B455} (1995)
  59--82
\href{http://www.arXiv.org/abs/hep-ph/9505244}{[{\tt hep-ph/9505244}]}.

\bibitem{Fischler:1981zk}
W.~Fischler, H.~P. Nilles, J.~Polchinski, S.~Raby, and L.~Susskind ``Vanishing
  renormalization of the {D} term in supersymmetric {U(1)} theories'' {\em
  Phys. Rev. Lett.} {\bf 47} (1981)
757.

\bibitem{Weinberg:2000cr}
S.~Weinberg ``The quantum theory of fields. {V}ol. 3: Supersymmetry''.
  Cambridge, UK: Univ. Pr. (2000) 419 p.

\bibitem{Fogleman:1983hm}
G.~Fogleman, G.~D. Starkman, and K.~S. Viswanathan ``Two loop calculation of
  the effective potential for the {W}ess-{Z}umino model'' {\em Phys. Lett.}
  {\bf B133} (1983)
393.

\bibitem{Fogleman:1984eh}
G.~Fogleman and K.~S. Viswanathan ``The effective potential for chiral
  supersymmetric models'' {\em Phys. Rev.} {\bf D30} (1984)
1364.

\bibitem{Miller:1983ci}
R.~D.~C. Miller ``The two loop effective potential of the {W}ess-{Z}umino
  model'' {\em Nucl. Phys.} {\bf B241} (1984)
535.

\bibitem{Davydychev:1992mt}
A.~I. Davydychev and J.~B. Tausk ``Two loop selfenergy diagrams with different
  masses and the momentum expansion'' {\em Nucl. Phys.} {\bf B397} (1993)
123--142.

\bibitem{Davydychev:1993pg}
A.~I. Davydychev, V.~A. Smirnov, and J.~B. Tausk ``Large momentum expansion of
  two loop selfenergy diagrams with arbitrary masses'' {\em Nucl. Phys.} {\bf
  B410} (1993) 325--342
\href{http://www.arXiv.org/abs/hep-ph/9307371}{[{\tt hep-ph/9307371}]}.

\bibitem{Misiak:1994zw}
M.~Misiak and M.~Munz ``Two loop mixing of dimension five flavor changing
  operators'' {\em Phys. Lett.} {\bf B344} (1995) 308--318
\href{http://www.arXiv.org/abs/hep-ph/9409454}{[{\tt hep-ph/9409454}]}.

\bibitem{Ford:1991hw}
C.~Ford and D.~R.~T. Jones ``The effective potential and the differential
  equations method for {F}eynman integrals'' {\em Phys. Lett.} {\bf B274}
  (1992)
409--414.

\bibitem{Ford:1992pn}
C.~Ford, I.~Jack, and D.~R.~T. Jones ``The {S}tandard {M}odel effective
  potential at two loops'' {\em Nucl. Phys.} {\bf B387} (1992) 373--390
\href{http://www.arXiv.org/abs/hep-ph/0111190}{[{\tt hep-ph/0111190}]}.

\bibitem{Martin:2001vx}
S.~P. Martin ``Two-loop effective potential for a general renormalizable theory
  and softly broken supersymmetry'' {\em Phys. Rev.} {\bf D65} (2002) 116003
\href{http://www.arXiv.org/abs/hep-ph/0111209}{[{\tt hep-ph/0111209}]}.

\end{thebibliography}\endgroup
